%% file: ms_ApJ86798_v3.tex
\documentclass[12pt,preprint2]{aastex}

\usepackage{natbib}
\usepackage{lscape,graphicx}

\shorttitle{Dynamic Evolution of Young Radio Sources}
\shortauthors{An and Baan}

\begin{document}

\def \kms {km s$^{-1}$ }
\def \lsol {L$_{\odot}$ }
\def \msol {M$_{\odot}$ }
\def \kmss {km s$^{-1}$}
\def \lsols {L$_{\odot}$}
\def \msols {M$_{\odot}$}

\title{The Dynamic Evolution of Young Extragalactic Radio Sources}

\author{Tao An\altaffilmark{1,2,3} and Willem A. Baan\altaffilmark{2}}
\altaffiltext{1}{Shanghai Astronomical Observatory, Chinese Academy of Sciences, 200030, Shanghai, China; antao@shao.ac.cn}
\altaffiltext{2}{ASTRON, P.O. Box 2, 7990 AA Dwingeloo, The Netherlands;  baan@astron.nl}
\altaffiltext{3}{Key Laboratory of Radio Astronomy, Chinese Academy of Sciences, Nanjing 210008, China}

\begin{abstract}
The evolution of symmetric extragalactic radio sources can be characterized by four distinct growth stages of the radio luminosity versus source size of the source.  The interaction of the jet with the ambient medium results in the formation and evolution of sources with non-standard (flaring) morphology. In addition, cessation or restarting of the jet power and obstruction of the jet will also result in distinct morphological structures. The radio source population may thus be classified in morphological types that indicate the prevailing physical processes.
Compact Symmetric Objects (CSOs) occupy the earliest evolutionary phase of  symmetric radio sources and their dynamical behavior is fundamental for any further evolution. Analysis of CSO dynamics is presented for a sample of 24 CSOs with known redshift and hotspot separation velocity and with a large range of radio power.
Observables such as radio power, separation between two hotspots, hotspot separation velocity and kinematic age of the source, are found to be generally consistent with the self-similar predictions for individual sources that reflect the varying density structure of the ambient ISM. Individual sources behave different from the group as a whole. The age and size statistics confirm that a large fraction of CSOs does not evolve into extended doubles.

\end{abstract}
\keywords{galaxies: active -- galaxies: jets -- galaxies: evolution}

\section{Introduction}

The morphological shape of all extragalactic radio sources is characterized by distinct structural components that may be more or less prominent. 
The jet-hotspot-lobe-cocoon structure of the largest classical double sources has first been classified in two distinct populations  \cite{FR74}.  FR\,I sources have lower radio powers and less well-confined outer lobes energize by a lossy jet and hotspots that are at less than 50 percent of the total extent of the source. Their outer structure becomes a flaring, meandering plume that is shaped by the motion of the galaxy or by internal fluid instabilities. A typical FR\,I source 3C 31 has a relatively prominent inner jet and filamentary plumes extending 300 kpc \cite{Lai08}. 
In contrast, FR\,II sources have higher radio luminosity and well-defined outer lobes that surround faint jets and very prominent hotspots at more than 50 percent of the source extent. FR\,II sources also exhibit extended cocoon structures in low-frequency interferometric images. An archetypal FR\,II source, Cygnus A exhibits two symmetric edge-brightened lobes spanning over 160 kpc, a prominent hotspot at the leading edge of each lobe, but with a weak flat spectrum core and intervening jets \cite{CB96}.  This FR classification reveals a morphological sequence that relates to the ability of a jet to transport and deposit momentum and energy at the leading edge of the lobe.

These large-scale Symmetric Objects (LSO) form an extended family together with Medium-sized Symmetric Objects (MSO) and the smallest Compact Symmetric Objects (CSOs), which represent the earliest development phase for all double sources. Scaling up in age and size from CSOs, the MSO group comprises of the GHz-Peaked Spectrum (GPS) sources \citep{Fanti90, ODea91} \citep[review][]{ODea98,Fanti09}, the Compact Steep-Spectrum (CSS) sources \citep{PeaW82} \citep[review][]{Fanti94,Fanti09}. A further morphological group of core-jet Blazar sources with jets pointing at the observer includes  BL\,Lac Objects, High Polarization Quasars and Optically Violent Variable Quasars.

The structural and spectral characteristics of radio sources are determined by the power of the source, the local environment of the host galaxy, and the evolutionary age. In this respect, the compact and supposedly young CSOs (and GPSs) are critical for understanding the rest of the radio source population. Their compactness has been attributed to two distinct scenarios: (1) the  {\it youth scenario} \cite[e.g.][]{Phi82,Fanti95} suggests that they are in an early evolutionary stage, and may continue to grow to Mpc-scale extended radio sources, and (2) the {\it frustration scenario} \citep[e.g.][]{vanBre84,ODea91, Car94,Car98} suggests that their growth is retarded (stagnated) by the dense Inter-stellar Medium (ISM) within the host galaxy. 
Both scenarios may apply for CSOs (and GPSs) as they relate directly to the power level and the duration of the nuclear activity. Young sources with persistent long-term nuclear activity will continue to grow and eventually become LSO 
doubles. Other young sources may stagnate because of intermittent behavior of the nuclear activity and the jet power at any time during their evolution \citep{RB97, Kun06}.

The complicated fluid and radiative behavior of extra-galactic  sources driven by the jet outflows has been modeled using self-similarity of the flow pattern as it evolves and moves away from the host galaxy into regions with lower ambient densities. Using such simplified models for well-defined doubles, the radio luminosity and spectral characteristics as a function of density gradient and the size of the source have been considered by a number of authors \cite{Car94, KA97, Car98, KB07}.  The advancing motion of the lobes and hotspots is fully determined by the momentum carried by the jets, the locations of shocks in the jets, hotspots and lobes, and the density structures of the ISM of the host galaxy and the Inter-galactic Medium (IGM). On the other hand, the morphological radio structure of FR\,I-like sources is very different from the structure of FR\,II-like sources. After the formation of a weak hotspot at a certain distance from the nuclear engine, the flow is conically confined and forms a flaring (and meandering) structure extending for a large distance, as modeled for the well-known FR\,I source 3C31 \citep[see][]{Wang09}.

The current investigation considers the observable properties of a well-defined sample of CSO sources, in order to determine their structural evolution with time. The observable properties of CSO are important to understand the early evolution of radio sources and how they evolve into the larger population of extra-galactic radio sources. 
Earlier modeling using self-similarity of the shape of the FR\,II-like radio sources assumed a constant separation velocity of the radio hotspot. However, analytic modeling of the expansion of hotspot and cocoon indicates that the radio sources experience a deceleration from CSO to MSO phases, and an acceleration process from MSO to LSO-FR\,II phases \citep{KK05,KK06}. While such a trend would also follow from simple dynamical analysis and from the analogies with fluid dynamics, the hotspot separation velocity will be one of the observables in the study the earliest CSO stage. 

The radio morphology of all members of the family of radio sources represents the evolutionary stage of the source but also reveal symptoms of survival or demise. In order to connect the early evolution of CSOs with the later evolution of the radio source family, the physical processes that determine the morphology and the radio properties will be used to further classify the evolutionary phenomenology. Seven morphological types within the CSO, MSO, LSO family are used to incorporate and describe what may happen to extra-galactic radio sources during their life time.

This paper presents the following components: a description of the dynamical evolution of extra-galactic radio sources and the place of CSO sources within a morphological classification scheme (Section 2), a description of the observational data and the CSO sample used for this work (Section 3), the observed dynamic changes of CSO sources (Section 4), the statistics of the CSO sample (Section 5), and conclusions (Section 6).  

\begin{figure*}[ht]
\center
\includegraphics[scale=0.80,angle=0]{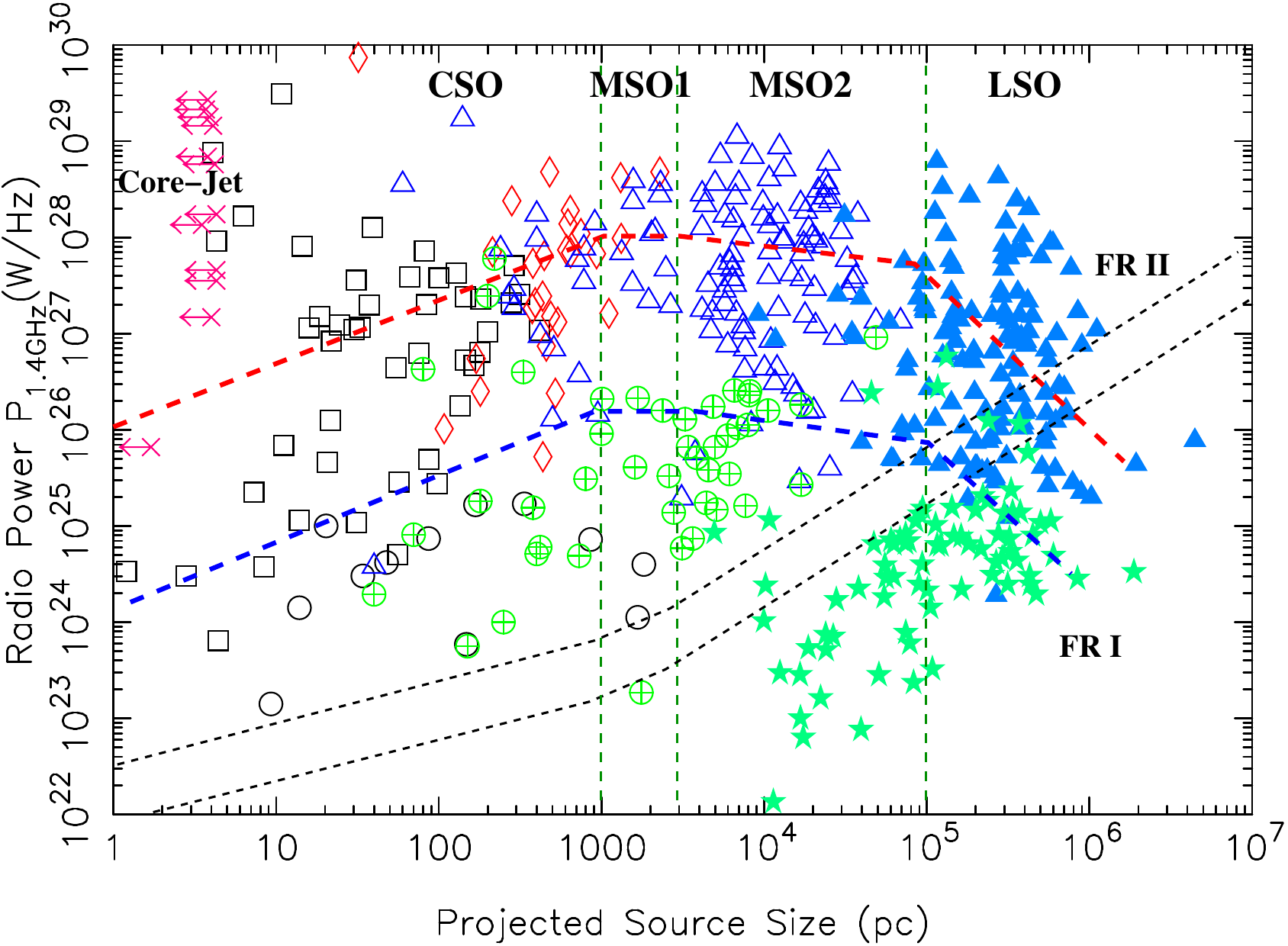}
\caption{
The radio power versus the linear extent of large scale radio sources: Compact, Medium-sized and Large Symmetric Objects. Exemplary evolutionary tracks based on parametric modeling are depicted for the high-radio-power and low-radio-power sources using red and blue dashed lines. The black dashed lines marks the (approximate) boundary region between stable laminar jet flows (above the lines) and unstable turbulent flows (below the lines). Symbols represent different morphological and spectral classes of radio sources (number of data points in bracket): black square: CSO (46), black circle: low-power GPS (12), red diamond: high-power GPS (27), purple cross: HFP (12), green circle: low-power CSS (114), blue open triangle: high-power CSS (44), blue filled triangle: FR\,II (140), and green filled star: FR\,I  (79). Catalog references for Figure 1: Akujor \& Garrington 1995 (High-Power CSS); Black et al. 1992 (FR II); Bogers et al. 1994 (FR II); Bridle et al. 1994 (FR II); Dallacasa
et al. 2002a (High-Power CSS); Dallacasa et al. 2002b (High-Power CSS); de Vries et al. 2009 (Low-Power GPS); Fanti et al. 1987 (FR I and FR II); Fanti et al. 2001
(High-Power CSS); Kunert-Bajraszewska et al. 2010 (Low-Power CSS; K10); Kunert-Bajraszewska et al. 2006 (Low-Power CSS; K06); Laing et al. 1983 (FR I and
FR II); Leahy et al. 1989 (FR II); Leahy \& Perley 1991 (FR I and FR II); Liu et al. 2007 (High-Power GPS; L07); Orienti \& Dallacasa 2012 (HFP; OD12); Orienti
et al. 2007 (HFP; O07); Orienti et al. 2010 (HFP; OD10); Orienti et al. 2006 (HFP; O06); Stanghellini et al. 1998 (High-Power GPS; S98) and Xiang et al. 2006
(High-Power GPS; L06).
(A color version of this figure is available in the online journal.)
\label{fig1}}
\end{figure*}

\section{Evolution of the Radio Structure}

The variation of the radio power $P_{rad}$ versus the total extent of the source $D$ for the whole family of extra-galactic radio sources serves to further understand their evolutionary characteristics (Fig. \ref{fig1}). Such diagrams have been presented earlier in the literature \citep[e.g.][]{ODeaB97, Kun10}. The current ($P_{rad}-D$) diagram contains similar but also different groups of data points than earlier versions, particularly for the region of early development of CSO radio sources. A sample of low-power GPS sources \cite{deV09} has been included in the bottom-left corner of the diagram, as likely precursors of the low-power MSO sources \cite{KB10}. In addition, there are more data points for the group of low-power FR\,Is ($<10^{24}$ W/Hz) \cite{LRL83, Fanti87, FPD01}.   

\subsection{Evolution of radio doubles}

The dynamic evolution of extra-galactic double radio sources is characterized by observables such as the kinetic power provided to the lobes, the total extent of the source, the advancing velocity of the terminal hotspots and the outer lobes, and depends on the density gradients of the ambient medium in the host galaxy along the path of the jets and the lobes. Modeling of the hotspots, the lobes, and the bow-shock has been done using self-similarity arguments to maintain the overall shape of a symmetric double source \cite[e.g.,][]{Beg89, Beg96, Car94, Car98, KA97, KB07}.

The radiative properties of the structural components are determined by the balance between the particle energy density, adiabatic losses resulting from expansion into lower pressures regions, and at later stages the evolving energy density of the magnetic fields in the hotspots and lobes, and the energy density of the Cosmic Microwave Background (CMB). 
The radiative properties of the hotspots and the well-confined lobes may thus be derived from modeling \cite{KA97, KK06, KB07}, although the assumption of self-similarity limits the interpretation of some observable parameters.

The density gradient along the path of the jet ({\it i.e.}, in the direction perpendicular to the galactic plane) plays a dominant role during the early evolution of radio sources. The general form used for this gradient is $\rho(z)$ = $\rho_{0}(a_{0}/z)^{\beta}$ \cite{Ki72}, where $\rho_{0}$ and $a_{0}$ are the reference density and the scale length of the ISM and the IGM, respectively. Observational data of elliptical galaxies, galaxy groups and clusters suggest a double-$\beta$ model \cite{XW00,FMO04} with an innermost atmosphere of roughly constant density and a steepening density profile beyond $a_0$. 
A transition at $a_{0} \approx1$ kpc from the galactic ISM ($\beta \approx 0$) to the IGM ($\beta \ge 1.5$) \cite{KB07} is confirmed by the observed discontinuity of the hotspot radius to arm length relation \cite{KNK09}. However, $a_{0}$ can become significantly larger for certain galaxies.

\subsection{Stages of evolution of radio doubles \label{evol}}

The existing literature provides a fundamental understanding about the dynamical evolution of FR\,II-like double radio sources. Although there is no full agreement about the actual parameterization of this evolution, four evolutionary radiative stages for a source with constant jet power $P_{src}$ may be identified that are related to the dominant physical processes \citep[see][]{KB07}: 
\newline {\bf Phase 1 - CSO} -  The radio luminosity evolution in this stage of CSOs and compact GPS sources is characterized by a flat density profile (up to a source size of about 1--3 kpc) with $\beta$ = 0.  During this first stage adiabatic losses in the hotspots and lobes dominate the energy dissipation process as the embedded B-fields are not yet strong enough to cause dominant synchrotron radiation losses. The radio power increases with time as $P_{rad} \propto t^{2/5}$ and with the source size as $P_{rad} \propto D^{(8-7\beta)/12} = D^{2/3}$ because of a steady increase in the conversion efficiency from jet kinetic power to radiative power. The spectrum should be steep with index $\alpha$ = 1.0.
\newline {\bf Phase 2 -  MSO-1} - This stage of extended GPS and more compact CSS sources occurs in the transition region ($a_{0} = 1-3$ kpc and $\beta \approx 1$) between the inner galaxy with a flat density profile and the outer galaxy with a steeper profile. During this phase there is a balance between adiabatic losses and synchrotron losses in the hotspots and lobes. The radio luminosity varies as $P_{rad} \propto D^{0}$. The radio spectrum is again rather steep with $\alpha$ = 1.0.
\newline {\bf Phase 3 - MSO-2} - The third stage is mostly occupied with CSS sources for which the synchrotron losses resulting from entrained B-fields dominate the adiabatic expansion losses resulting from the steep density gradient ($a_{0} \ge 3$ kpc and $\beta \ge 1.5$) of the external (inter-)galactic environment.   The radio luminosity decreases steadily with source size as $P_{rad} \propto D^{(8-7\beta)/12} = D^{-0.2}$, while the radio spectrum would be relatively flat with $\alpha$ = 0.5. 
\newline {\bf Phase 4 - LSO} - The fourth evolutionary stage is populated with fully developed FR\,II and FR\,I sources. The density profile of the IGM follows a power law with $\beta \ge 1.5$. The radio luminosity decreases sharply as $P_{rad} \propto D^{(-4-\beta)/(5-\beta)}=D^{-1.6}$ and the radio spectra are steep with $\alpha$ = 1.0. This phase may start around 100 kpc, where the inverse Compton losses resulting from the CMB energy density dominate over the synchrotron losses.

The above described evolutionary stages have been depicted in the luminosity - size ($P_{rad} - D$) diagram of Figure \ref{fig1} for two representative power levels, that represent a complete CSO--MSO--LSO luminosity evolution. Along these evolutionary tracks it is  assumed that the jet power $P_{src}$ remains constant during the whole evolutionary lifetime until the LSO FR\,II stage is reached. The shape of this radio power $P_{rad}$ track is fully determined by the ($\beta - D$) relation.

 \subsection{Jet stability and flaring sources \label{stability}}

The $P_{rad}-D$ evolution of FR\,I-like flaring sources can partially coincide with that of the well-defined FR\,II-like double;sources. During the flaring stage the self-similar modeling for the morphology and the radiative properties would apply. Although both may coexist in the same region of the ($P_{rad}-D$) diagram, the (downward) evolutionary tracks of flaring sources are determined by the lower radio power and the cooling of the expanding radio structures. 

The stability of the jet flow and the interaction with the ambient material determine the efficiency of the energy and momentum transport to the lobes as well as the relative radio prominence of the jets and the lobes.  After the oblique (reconvening) shock at the base of the jet converts the expanding flow profile to a nearly parallel (cylindrical) flow, a relativistic flow may persist over a long distance until eventually a decelerating shock forms at the entry of the lobe. High-power jets with laminar flow are most efficient for transporting the energy and momentum, which makes the jet mostly invisible in the radio maps of most prominent FR\,II-like doubles.

A fluid boundary layer exists between the supersonic and relativistic core of the jet flow and the stationary ambient medium that forms the velocity gradient from maximum to minimum velocity. This boundary layer dissipates kinetic energy because of viscous entrainment and fluid instabilities. 
With distance traveled the boundary layer grows in thickness and at a certain characteristic scale length after the confining shock this initially laminar boundary layer becomes turbulent, which further increases the entrainment and the momentum loss and weakens the jet. Eventually the boundary layer occupies the whole width of the jet. 
Under normal conditions, the supersonic jet with boundary layer forms a stagnation standing shock in the lobe \cite{KA97,Wang11}, which defines the location of the hotspot.  However, a jet with a boundary layer reaching its centerline has lost much of its energy and momentum and fails to form a standing shock at a stagnation point. This jet ceases to supply sufficient energy and momentum to the lobe, which becomes diffuse and undefined as in FR\,I sources \citep{Wang09}.

Stability analysis of the jet flow within the ambient medium predicts the distance at which the boundary layer has grown to cover the whole width of the jet \citep{KA97,Wang08, Wang11}. This distance may be expressed as:

$z_{e} = c_{1} D^{(\beta+4)/6}$ (1)

\noindent where the proportionality constant varies with the jet power and the properties of the density profile \citep{KA97}. This distance forms a specific boundary in the $P_{rad}-D$ diagram, beyond which the jets become ineffective in supplying energy and momentum to the lobes. 

Noting that the shape of this boundary in the $P_{rad}-D$ diagram (as a function of jet travel distance) remains model dependent, a representative boundary region has been entered in Figure \ref{fig1} (denoted by two black dotted lines) using the same $\beta-D$ variation as used for the radio luminosity tracks.  The vertical scaling of this region, which varies with the environmental parameters and the jet power, is not well-determined and has been chosen to incorporate some prominent FR\,I sources (3C31 with $z_e$ = 3.5 kpc and $P_{rad}$ = 2.0e24 W/Hz and  3C449 with $z_e$ = 17 kpc and $P_{rad}$ = 1.12e24 W/Hz) and matches the FR\,II - FR\,I boundary at larger values of $D$. Evolutionary tracks reaching into this stability region predict jet stagnation/disruption and flaring FR\,I-like lobe structures. In addition, this curve predicts that low-power CSO and MSO sources, or any source that experiences a systematic decrease (or cessation) of jet power, could directly evolve into a stagnated FR\,I-like source.

\subsection{Jet power continuity \label{jetpower}}

The continued evolution of radio sources described above depend primarily on the continuity of the nuclear activity and the persistence of the jet. 
Theoretical studies indicate that short-lived compact radio sources and the intermittent activity of the central engine may be caused by a radiation pressure instability within an accretion disk \citep{Czerny09}. According to these authors, a radio source powered by a short-lived outburst of the central activity is not able to escape from the host galaxy unless the active phase lasts longer than $10^4$ year, which is typically the time required for the lobe to successfully pass the ISM-IGM boundary (see discussion in sections 2.3 and 4.6). 
The discontinuity of jet activity has been associated with the formation of FR\,I-like sources from CSSs because of intermittent and episodic AGN activity  \citep{Mar03, Kun06} and the death of CSO sources before reaching the MSO stage \citep{Sne99,Gug05,Kun10}.

A reduction of the jet power results in the formation of a hotspot receding from the leading edge (and the entry point) of the lobe, the jet becomes more lossy because of increased prominence of the jet boundary, and the lobe develops an instability-driven (meandering) structure as in FR\,I galaxies.   The emission spectrum of the lobe will steepen and the jet may develop surface instabilities.   Further reduction or termination of the jet power turns the source into a low-power relic as it eventually dies in its current phase. 

Recurrent  jet activity will result in various FR\,I-like morphologies depending on the jet power and the time separation between subsequent events. During a low-power interval, the lobe becomes more diffuse, the hotspot separation will also be retarded and the source moves downward in the $P_{rad}-D$ diagram. After re-ignition the retarded hotspot could again move to the entry of the lobe. A powerful recurrent source will expand its jet into the excavated channel of past activity and may exhibit a double-double morphology. Alternatively, restart in the jet along a different axis may result in X-shaped source. A long time separation between subsequent events (compared with the cooling lifetime) will make the source reappear as a new startup FR\,II-like source.

\subsection{The size and velocity of the hotspots \label{hotspots}}

The size of the hotspot results from a balance between the energy carried by the jet and losses from radiation and adiabatic expansion. The parametric solutions use a simplified assumption that the separation velocity of the hotspot remains constant during the early evolution \cite{Fanti95, Rea96, Car98,KA97}. However, constant-velocity models may not be representative for observed source behaviour. Analysis of the hotspot size versus $D$ relation shows that the observed hotspot size for CSOs has a different growth rate with arm length (as $D^{1.4}$ for $D<1$kpc) than the larger MSO (GPS, CSS) and LSO-FR\,II sources (as $D^{0.4}$ for $D>1$kpc), which also confirms the location of the ISM$-$IGM transition \citep{KK06, KNK09}. Because the change in size of the hotspot and the hotspot separation velocity are parametrically related, these rates suggests that the hotspot of CSOs should decelerate as $v_{HS} \propto D^{-1}$ (for $\beta$ = 0) and accelerate as $v_{HS} \propto D^{0.3}$ (for $\beta$ = 1.5) during later phases.    Under our simplified assumptions the $v_{HS }$ for the CSOs would decrease as $D^{-2/3}$ during Phase 1 with constant $P_{src}$ (Section \ref{model}). The variation in the detected surface brightness levels for different observing arrays prevents a test of this relation using the CSO data in this paper.

It should also be recognized that the jet remains at least supersonic in order to generate a hotspot. If $v_{HS}$ continues to decrease linearly until it reaches the ISM$-$IGM boundary, the initial hotspot advancing velocity should at least be 0.3$-$0.5 $c$  \citep[see][]{KK06,KNK09}.  Jet flows with velocities below this threshold become subsonic (of order $10^{-3} c$) before reaching the ISM$-$IGM boundary and these will die prematurely (see section 4.4). 

\subsection{Morphological Classification of Sources}

The physical processes that determine the evolution of radio sources create a large variety in the radio morphological structure of radio sources. The following morphological types (Mtype) may be assigned to members of the family of extragalactic radio sources. 

{\bf Mtype\,2 double sources} have energetic laminar jets with prominent well-confined lobes but weak or invisible central core emission and weak or no intervening jet structure. They have well-confined lobes and prominent terminal hotspots at the leading edge of the lobe with a sharp edge-brightened morphology, that is indicative of a strong advancing shock. A low hotspot size$-$to$-$linear extent ratio indicates dominant radial expansion. 
Mtype\,2 sources have sufficient and long-lived jet power and occur among the CSO, MSO, and LSO populations. Mtype \,2 sources  evolve along the evolutionary tracks in the radio power-size ($P_{rad}-D$) diagram.  FR\,II sources are the largest LSO sources within the Mtype\,2 double population.

{\bf Mtype\,1 flaring sources} are characterized by lossy prominent jets, diffuse, diverging, and less confined lobes or flaring lobes, and hotspots located away from the lobes. Mtype\,1 sources have jets where the boundary layer reaches the centerline (see section \ref{stability}), which results in a low-momentum, flaring and/or meandering flow constituting the radio lobe. 
Mtype\,1 sources exist among lower-power CSO, MSO and LSO sources that have reached the jet (in-)stability region in the P-D diagram (see Figure 1). Mtype\,1 MSO and LSO sources are the FR\,I sources with hotspots at $<$ 50\% of the arm lengths.

{\bf Mtype\,3 dying sources} experience cessation or reduction of the jet power, which result in the disappearance of the hotspots. The remaining radio structure continue its adiabatic expansion without the necessary energizing of the lobes. The sources becomes an expanding diffuse relic unless there is reactivation of the nuclear activity and re-formation of the jet.
Among CSOs the short-lived central source activity results in flow instabilities before the ISM$-$IGM transition point, and the eventual dissipation of the lobe structure. MSO and LSO sources will retain their shape without a hotspot and continue to expand while radiating their energy away.

{\bf Mtype\,4 restart sources} experience a restart of nuclear activity or intermittent activity and the startup of jet power into the earlier excavated channels. Depending on the new jet power, the hotspot advances through the channel until it reaches the lobe entry and sets up a new shock-confined lobe within a larger existing lobe. Depending on the time interval between the periods of activity, the new lobe may be quite distinct from the relic lobe structure, as in double-double radio sources. Regular intermittent activity may results in a series of shock fronts passing through the lobe. X-shaped radio sources could also signify a restart of the central source with an different jet axis.

{\bf Mtype\,5 obstructed sources} have lower-power jets in the CSO Phase\,I and MSO Phase\,II that interact with the surrounding medium along its path, which results in hotspots and misaligned outflows \cite{LB86}, in analogy to the {\it dentist's drill model} \cite{Sch82} or the {\it frustration model}, and significant sideways motions dissipating kinetic energy.   A higher hotspot size$-$to$-$linear extent ratio implies significant sideways expansion/motions. The morphology of these sources will deviate from the straight two-lobe picture to be expected from Mtype\,2.

{\bf Mtype\,6 core-jet sources} have a small angle between the jet-direction and the line of sight. These sources appear as one-sided jet with enhanced radio brightness because of doppler boosting.

{\bf Mtype\,7 trail sources} are trail radio galaxies where the Mtype\,1 lobe structures outside the ISM$-$IGM boundary are bent because of ram pressure resulting from the motion of the host galaxy throughout the intra-cluster medium.

\begin{figure*}[ht]
\centering
\includegraphics[scale=0.4,angle=0]{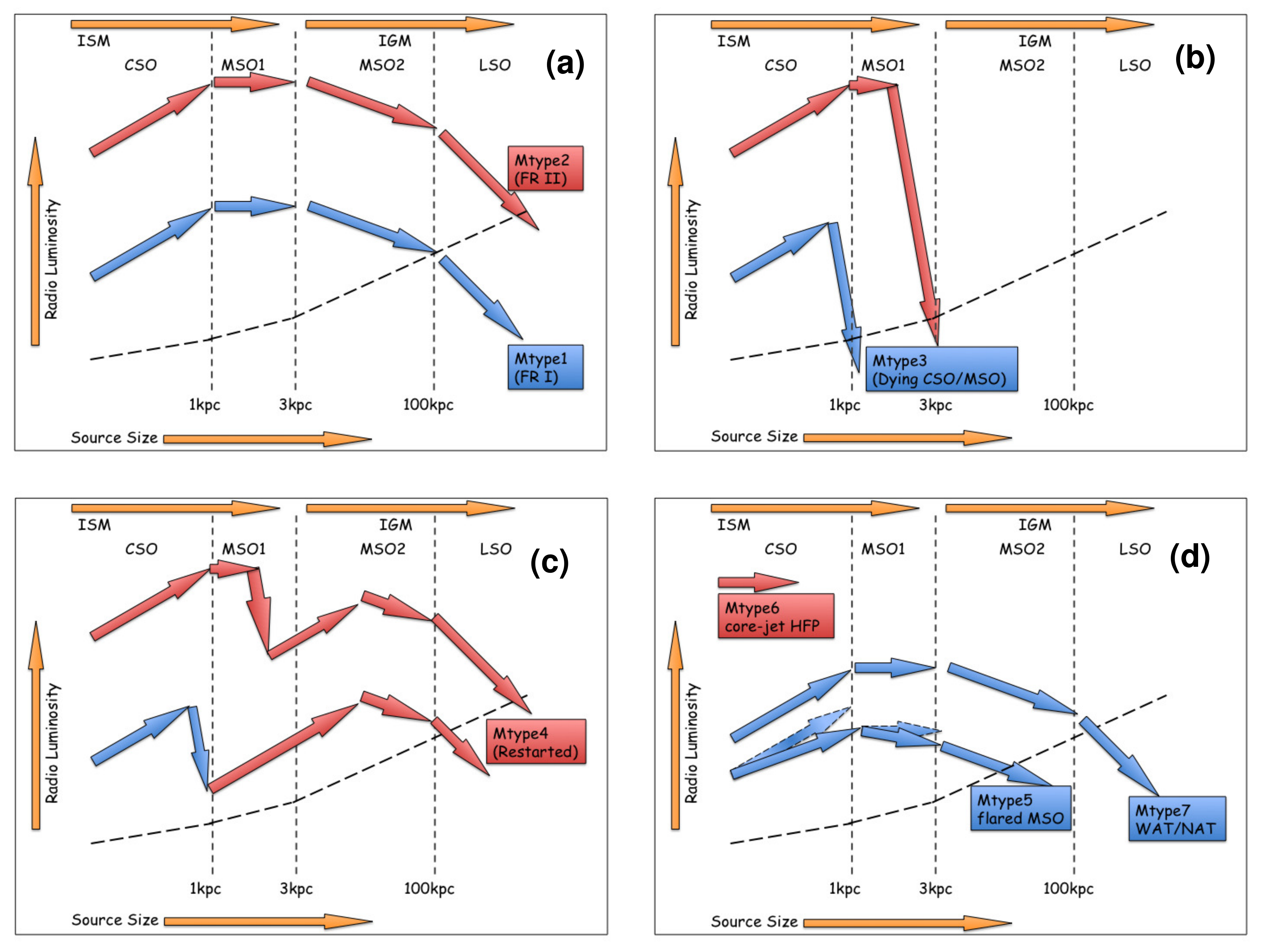}
\caption{Evolutionary pathways for extragalactic radio sources. Mtype\,2 double sources will continue to grow in size and move horizontally across the diagram (panel a: red  arrow). Low-power Mtype\,1 double sources continuouly grow till a distance of about 100 kpc where the jet becomes unstable and flaring. Mtype\,3 dying sources with decreasing or terminating power will move downward in the diagram and end up with relics (panel b). Re-energized and intermittent Mtype\,4 re-started sources will move upward into the standard evolutionary tracks ( (panel c). Some low-power CSOs may evolve into Mtype \,5 sources whose characteristic morphology is a hotpot within 1 kpc and a plume-like diffuse structure beyond the hotspot (panel d). Mtype \,6 sources are shown as core-jet objects occupying the top-left corner in the P-D diagram.  Some Mtype \,1 sources, which are shaped by the ram pressure of the IGM, are classified as Mtype \,7 (panel d). All sources below the jet stability line are not able to evolve into Mtype\,2 double sources (panel a). 
\label{fig2}}
\end{figure*}

\subsection{Evolutionary scenarios}

The long-term evolutionary tracks through the $P_{rad}-D$ diagram depicted in Figure \ref{fig1} reflect the long-term Mtype\,2 CSO-MSO-LSO double source evolution. All double radio sources will start out on such a track and they will continue to follow it as long as their jet power remains at the appropriate level.
These evolutionary tracks also indicate that the Mtype\,2 FR\,II double sources will naturally evolve into Mtype\,1 flaring FR\,I sources, because they eventually reach the region of jet instability (see Section \ref{stability}). This evolutionary path from FR\,II to FR\,I has been depicted graphically (blue arrows) in Figure \ref{fig2}.

Cessation or changes of the jet power could happen at any moment during the CSO-MSO-LSO sequence.
Mtype\,3 dying sources undergoing a lowering or cessation of the jet power move downward (red dashed arrows) in the $P_{rad}-D$ diagram and may be found among the Mtype\,2 double sources.
Mtype\,4 restart sources experiencing an increase in power will move upward in the diagram (black dashed arrows) to a higher power Mtype\,2 source. Re-energizing a Mtype\,3 source may again turn the source into a Mtype\,2 double.

Mtype\,5 obstructed sources with some signs of disruption may be found throughout the whole CSO-MSO region in the $P_{rad}-D$ diagram. 

All sources having reached the jet stability  region are already destined to become Mtype\,1. 

The scenarios for radio source evolution presented here are distinctly different from those presented by  \cite{Kun10}. These authors focus on the evolution from CSS to FR\,II and FR\,I and do not address the earliest evolutionary CSO stages considered in this paper. 
While the long-term evolutionary scenario for Mtype\,2 sources is similar for the scenario in Kunert-Bajraszewska et al. (2010), the details and the origin of Mtype\,1 sources is different. Because Mtype\,3 sources occur naturally in the whole diagram, the formation of low-power MSO-CSS sources does not (only) have to result from multiple interactions of a high-power CSS with obstacles in the host galaxy. Intrinsically low-power CSOs naturally evolve into low-power CSSs and also a reduction of the jet power in a CSO will result in a low-power CSS.

\section{The CSO data sample}

In order to compose a more complete sample of the CSOs for the purpose of a kinematics study, observations have been made of a sample of 10 CSO candidates without reliable proper motion measurements. The new observations were carried out at five frequency bands ranging from 1.6 to 15 GHz with the Very Long Baseline Array in 2005, and with a combined array of the Chinese VLBI Network (CVN) \citep{Li10} and the European VLBI Network (EVN) in 2009. The newly obtained data combined with the archival VLBI data made it possible to determine the separation velocities of the hotspots or to set upper limits for all ten sources \cite{An12}. Six of these sources are identified as CSOs and one as a CSO candidate, which increases the sample of CSOs and candidates with known proper motions from 30 to 37. 

The classification of CSOs is based on the morphology and spectral index criteria. A CSO is defined as a compact source (overall size $<$1 kpc) with two extended emission components on either side of a central flat-spectrum core. While the central core may be too weak for VLBI imaging, the presence of two extended components with steep spectra and edge-brightened morphology would be enough for CSO classification.  The mirror symmetry of the proper motions of the hotspots and internal jet knots with respect to the geometric centre provides supplementary CSO identification. Core-jet sources have not been included in the sample, because their radio luminosity is affected by an (unknown) Doppler boosting factor. A known redshift is required to convert from projected angular size to physical linear size. 

A sample of 46 CSOs with known redshift and size information has been presented in Table 1. A total of 24 sources in this sample also have published CSO proper motion data for the purpose of a statistical study of the kinematic properties of CSOs. Table 1 contains the following entries:

Column (1) gives the object name. 

Columns (2) and (3) present the redshift $z$ and luminosity distance $D_L$, which are taken from the NASA Extragalactic Database (NED) and from the literatures. 

Columns (4) and (5) present the overall source size in both angular size ($\theta_{AS}$) and projected linear size ($LS$). The calculation of $D_L$ and the conversion from $\theta_{AS}$ to $LS$ make use of a cosmological model with H$_0$ = 73 km s$^{-1}$ Mpc$^{-1}$, $\Omega_M$ = 0.27, and $\Omega_\Lambda$ = 0.73.

Column (6) gives the angular separation velocity of the hotspot $\mu={\theta_{AS}}/{\Delta t}$, which is calculated as the rate of the separation of two hotspots over the time interval. For most sources, $\mu$ is determined as a relative separation velocity with one hotspot as the reference. For CSOs with a visible core, the proper motions of hotspots are calculated as the separation velocity of each hotspot with respect to the central core. 
When multiple measurements are available, the value with the least uncertainty has been used.  

Column (7) gives the apparent transverse velocity in the source rest frame $v$ in units of $c$ obtained from $\mu$: $v=0.0158 \mu D_A (1+z)$, where $D_A$ is the angular size distance of the source in Mpc and $\mu$ is the angular separation velocity in mas yr$^{-1}$. 

Column (8) presents the kinematical age in the source rest-frame, ${\theta_{AS}}/{\mu (1+z)}$ or $LS/v$. 

Column (9) and (10) list the morphology classification based on the VLBI images and the reference for each object  (see details of the morphology classification in Section 2.5).  

Columns (11) to (13) present the flux densities $S$ at 1.4, 4.8 and 8.4 GHz, respectively. 

Column (14) presents the calculated turnover frequency.

Column (15) presents the spectral index $\alpha^{8.4}_{4.8}$ between 4.8 and 8.4 GHz (defined as $S_\nu \propto \nu^{-\alpha}$). A sample of seven CSOs were monitored using the VLA at 8.5 GHz and they showed little variability (mean rms of 0.7\%) over a period of eight months \cite{FT01}. Other CSOs, such as OQ~208, display much larger flux variations\cite{deB90,Sta97,Wu12}. Quasi-simultaneous multiple-frequency flux density measurements of CSOs  have been used preferentially from two available datasets: VLA A-array data at 0.3, 1.4, 4.8 and 8.4 GHz at three epochs during 1984 and 1991 \cite{Sta98}, and VLA data at six frequency bands ranging from 1.4 to 43 GHz in 2003-2004 \cite{Ori07}. For CSOs without simultaneous data, flux densities have been used from the Green Bank Telescope at 1.4 and 4.85 GHz \cite{WB92, GC91} and the VLA-A at 8.4 GHz \cite{Pat92,Hea07}. 

Column (16) lists the absorption-corrected radio power at 1.4 GHz in units of W Hz$^{-1}$ determined as $P_{1.4GHz}=4\pi D^2_{L} (1+z)^{(\alpha-1)} S_{1.4GHz}$. CSOs often exhibit inverted spectra with a turnover at a few GHz \cite{Sta98,Ori07} caused by synchrotron self-absorption and/or free-free absorption (FFA). The observed flux densities in Columns (11)--(13) show an $S_{1.4GHz}$ much lower than the 1.4-GHz flux density extrapolated  from $S_{4.8GHz}$ and the spectral index $\alpha^{8.4}_{4.8}$, which is in agreement with a spectrum turnover at frequencies higher than 1.4 GHz (Column 14).   Since $\alpha^{8.4}_{4.8}$ can be a reasonable representation of the optically-thin section of the spectrum, the 4.8-GHz flux density ($S_{4.8GHz}$) and spectral index $\alpha^{8.4}_{4.8}$ have been used to extrapolate the absorption-corrected 1.4-GHz flux density used for computing $P_{1.4GHz}$. Those sources with a turnover much higher than 5 GHz, the 8.4-GHz flux density and the spectral index between 8.4 and 15 GHz have been used \cite{Sta98,Ori07}.

\section{Dynamic Evolution of CSOs}

The smallest and youngest members of the radio source population are the CSO sources with linear sizes less than 1 kpc. All radio source must go through this stage where they are still submerged in the dense nuclear ISM of the host galaxy, which provides the strongest resistance against growth. The CSO sources occupy the left side of the radio power-size $P_{rad}-D$ diagram (Fig. 1) and show a range of five orders of magnitude in radio power. Because the adiabatic losses dominate over the synchrotron losses for these compact sources, the $P_{src}$ remains larger than $P_{rad}$ until the ISM$-$IGM boundary radius $a_0$. The morphological type to be expected during this early evolutionary stage ranges between Mtype\,2 doubles, Mtype\,1 flaring sources at the lower end of the $P_{rad}$-range, short-lived Mtype\,3 dying sources, and Mtype\,5 obstructed sources.The dynamical properties of the CSO sample will thus provide essential information about the evolutionary stage of the sources.

The observable parameters that may be used to study the dynamic properties of the CSO sources are (1) the hotspot separation velocity ($v_{HS}$ equals twice the hotspot advance velocity) assuming the jets are ejected with the same velocity on opposite sides, (2) the overall linear size of the source ($D$), and (3) the radio power ($P_{rad}$), which will equal the total jet power $P_{src}$ at the ISM$-$IGM transition distance. The dynamic behavior of the CSO sources are subject to variation of the jet kinematic and radiative powers, the variation in the ISM and IGM density gradients, and the initial density as well as scale height.
Projection of the source structure on the plane of the sky will lower the observed values of the projected source size and the hotspot separation velocity. The radio power and the kinematic age of the source are not affected by projection.

\subsection{Modeling the CSO stage \label{model}}

The evolution of the lobes and the hotspots of sources during the CSO stage have been modeled assuming self-similarity and expansion into a fixed opening angle. Assuming that $P_{src}$ is constant during the CSO phase, this results in an overall momentum balance equation at the bow shock \cite{KA97}: 

$P_{src} = c_{2} D^{2-\beta} v_{HS}^{3}$ (2)

\noindent where $P_{src}$ is the source mechanical power and $v_{HS}$ is the hotspot separation velocity, assuming that the this velocity varies proportionally with the bow shock and lobe separation velocities. In addition, the linear extent $D$ varies with evolutionary time $t$ as \cite{KB07}:

$D = c_{3} (P_{src} /(\rho a_{0}^{\beta}))^{\beta - 5} t^{3/(5 - \beta)}$ (3)

\noindent During the CSO stage, the radio power is less than $P_{src}$ and varies as \cite{KB07}:

$P_{rad} = c_{4} D^{(8 - 7\beta)/12}$ (4)

\noindent The proportionality constants in these equations depend on the properties of the ISM and the power of the source as presented in the above references.

The above equations with $\beta = 0$ may be used to describe the dependences of the observables presented in this paper, these being the distance of the lobe $D$, the hotspot (or lobe) separation velocity $v_{HS}$, the observed radio power $P_{rad}$, and the evolutionary time $t$ measured as 'kinematic age''.  

\subsection{Radio Power versus Linear Size} 

\begin{figure}[t]
\includegraphics[scale=0.23,angle=0]{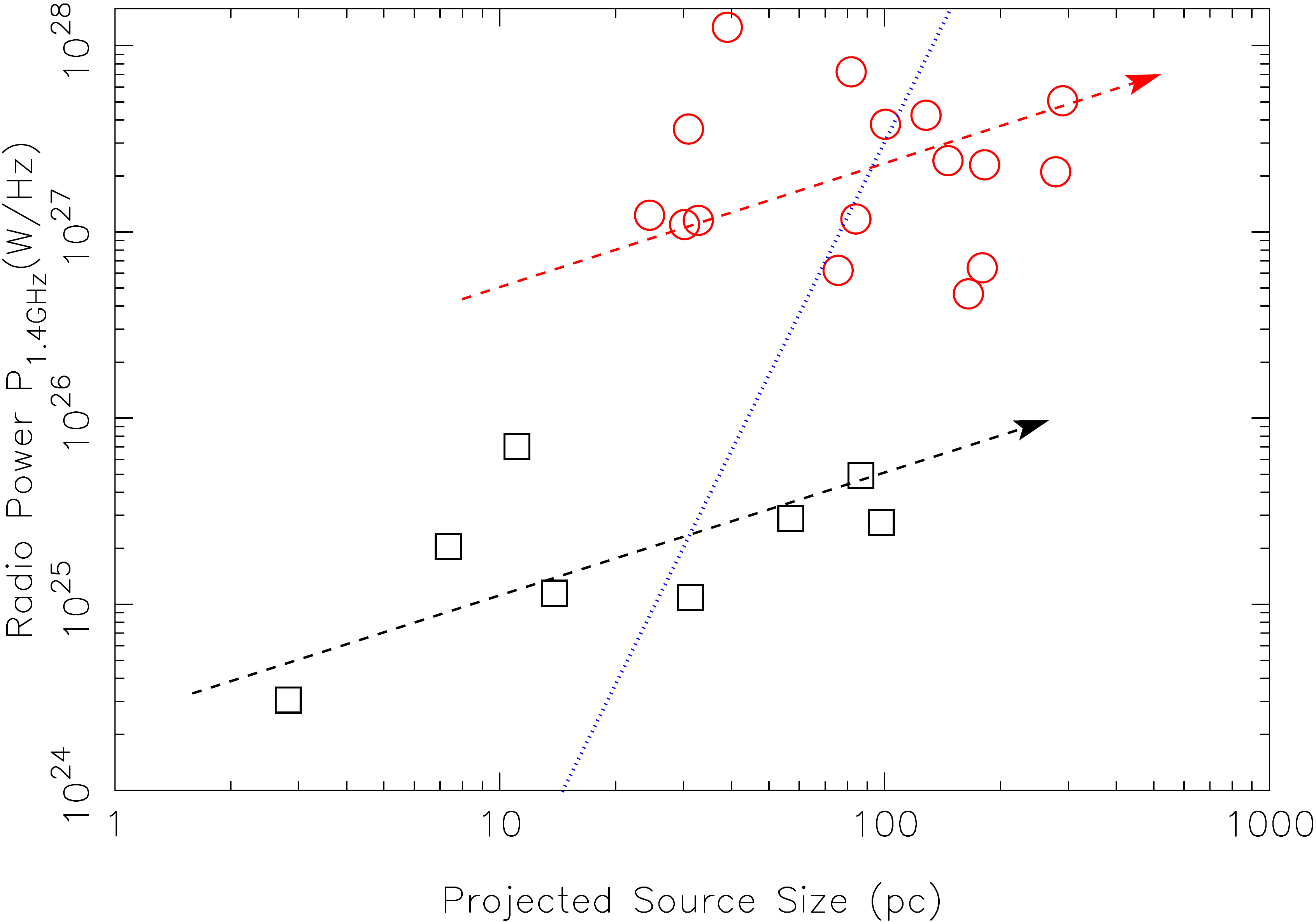}
\caption{Variation of the CSO radio power with projected linear source size. Symbols indicate high (red circles) and low (black squares) radio powers. Sources with constant jet power $P_{src}$ have a predicted evolution $P_{rad} \propto  D^{2/3}$ indicated by the dashed arrows.  The general trend in the distribution is represented by a $P_{rad}\propto D^{4}$ relation (dotted line). 
\label{fig3}}
\end{figure}

More powerful sources should be larger for a certain evolutionary age. However, there is no uniform relation between radio power and source size. The modeling results (Section \ref{model}) suggests a  $P_{rad} \propto D^{2/3}$ relation depicted as dashed arrows in Figure \ref{fig3}. This relation reflects the increase of the conversion efficiency from the kinetic power of the jet to the radio power as sources get larger and synchrotron losses increase. The spread in the data follows from variation of the source power $P_{src}$ and the ISM density structure. Projection effects would shrink the apparent source size and move a source to the left of the diagram.

More powerful sources can indeed be larger but they may also be older. The relation between the source size and kinematic age in Figure \ref{fig7}-b confirms that older sources are in general larger in size. The lower power sources (squares) reveal this trend in Figure \ref{fig3}. High-power sources (circles) show a different distribution, possibly resulting from source selection effects, but they may still behave according to the same relation.

The radio power is the measured quantity, while the physics of the source is determined by the source power $P_{src}$. Because $P_{rad}$ would equal $P_{src}$ at distance $D$ = $a_{0}$ = 1 kpc, an extrapolation using $P_{rad} \propto D^{2/3}$ to this point will provide an estimate for the source power $P_{src}$. 

\begin{figure}[t]
\includegraphics[scale=0.23,angle=0]{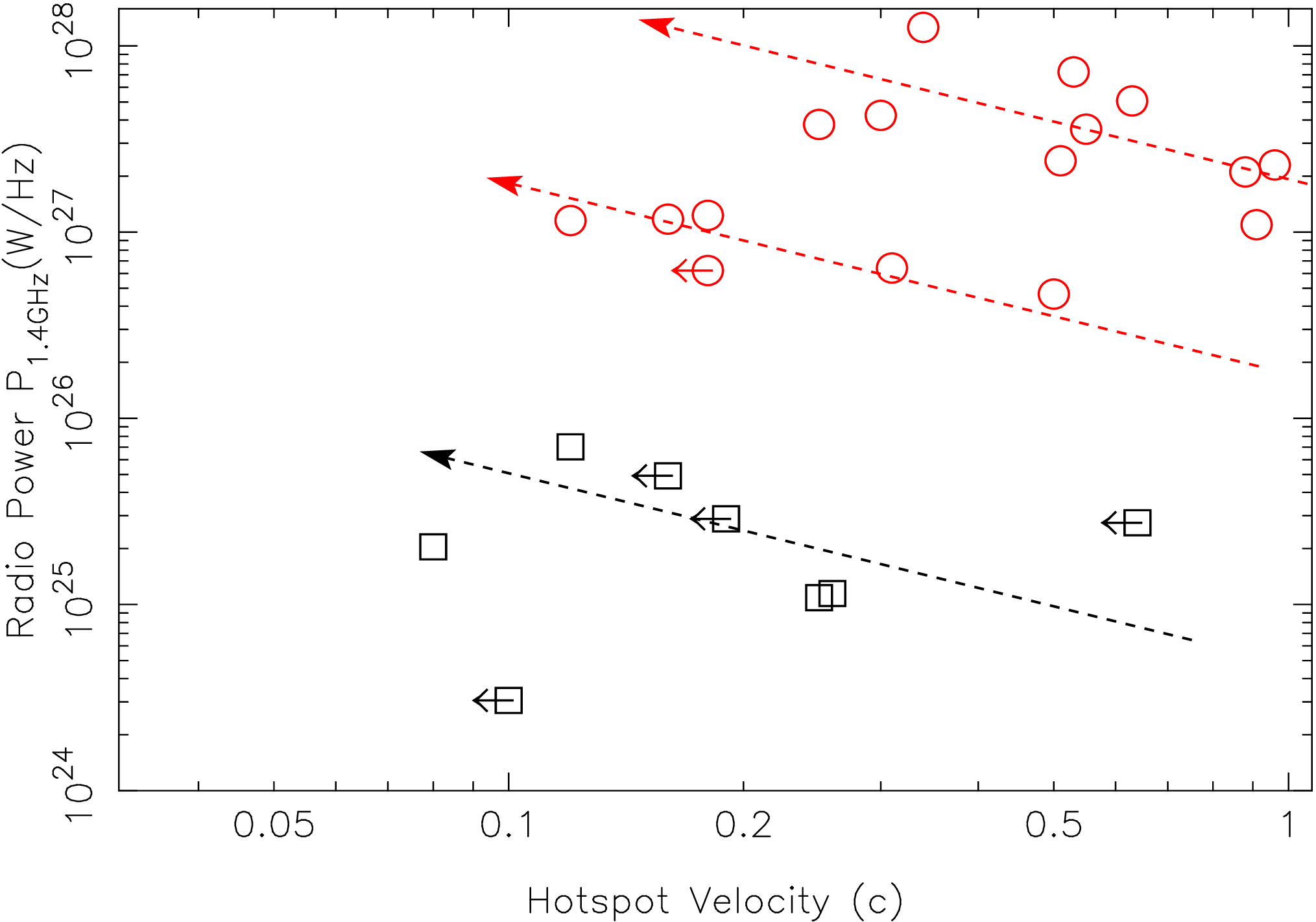}
\caption{Variation of the CSO radio power versus the hotspot separation velocity. As a group, the sources with higher $v_{HS}$ generally have a higher radio power $P_{rad}$ and also a higher source power $P_{src}$. Individual sources are predicted to evolve from high-$v_{HS}$ and low-$P_{rad}$ to low-$v_{HS}$ and high-$P_{rad}$ with time as $P_{rad}\propto v_{HS}^{-1}$. Symbols indicate high-power (circles) and low-power (squares) CSOs. Arrows on the symbols indicate upper limits.   
\label{fig4}}
\end{figure}

\subsection{Radio Power versus Hotspot Velocity} 

More powerful sources should have larger hotspot separation velocities for a given environment and a given age. However, the diverse environments and ages of the CSO sample cause a spread in velocity in Figure \ref{fig4}. A simple relation following from momentum conservation and the variation of $P_{rad}$ with $D$ (Sec. \ref{model}) suggests that a source with fixed $P_{src}$ varies as $P_{rad} \propto v_{HS}^{-1}$, i.e. a source with a given $P_{src}$ will evolve to larger $P_{rad}$ and smaller $v_{HS}$ until it reaches the ISM$-$IGM transition point. 
As a group, the CSOs with higher $v_{HS}$ also have higher $P_{rad}$, because of their higher source powers $P_{src}$.  

\begin{figure}[t]
\includegraphics[scale=0.23,angle=0]{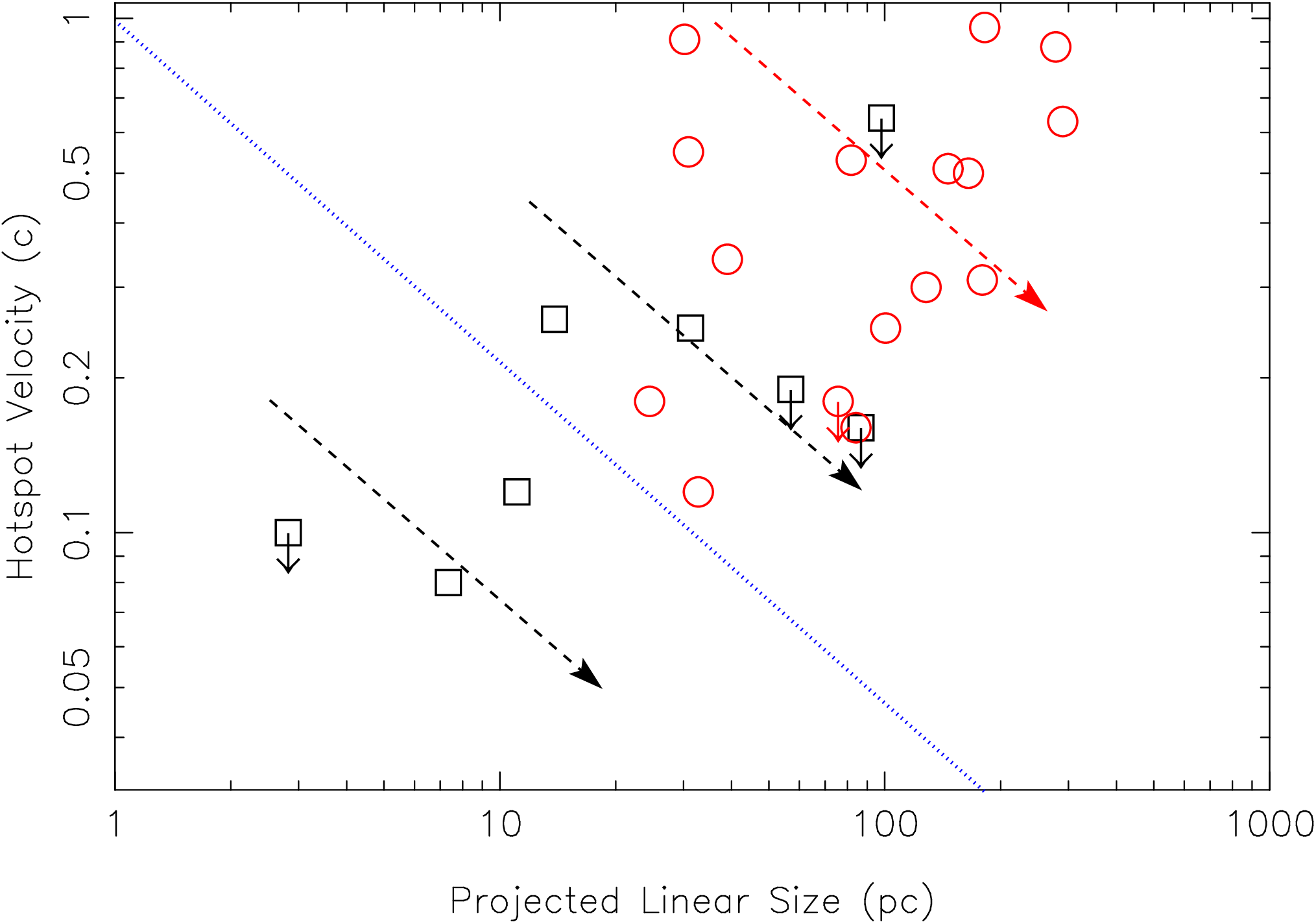}
\caption{Variation of the hotspot separation velocity versus the projected linear size. The dashed arrows $v_{HS} \propto D^{-2/3}$ indicate the predicted evolution for a source with constant power. The dotted line represents the lower boundary for the jet flow to continue to accelerate at the ISM$-$IGM transition point. CSO close to and below this line will not survive as double sources beyond the ISM. Projection effects for the CSOs shift data points diagonally towards the origin. 
\label{fig5}}
\end{figure}

\subsection{Hotspot Velocity versus Linear Size}  

The dynamics of the evolution of the radio source within the varying ambient environment should be well depicted by the variation of hotspot separation velocity with projected linear size. The $v_{HS}-D$ variation should be indicative of the validity of the parametric similarity solutions.

Modeling predicts that the hotspot separation velocity should vary as $v_{HS} \propto D^{-2/3}$ for a source with constant source power $P_{src}$ (Fig. \ref{fig5}). The separation velocity of an individual source expectedly decreases as the source becomes larger.  The $D^{-1}$ dependence derived from the variation of the  hotspot size is also consistent with these data \citep{KK06, KNK09}. As a group, the more powerful sources with faster moving hotspots will also be larger. This property may be represented by a $v_{HS} \propto D^{2/3}$ dependence.

As discussed in Section \ref{stability}, for a radio source evolving along the primary evolution track from CSO to CSS and to FR\,II stage, the hotspot separation velocity must at least be supersonic at the transition ISM$-$IGM transition point \cite{KK06,KNK09}. This means that a (survivor) CSO should lie above the (dotted line) $v_{HS} \propto D^{-2/3}$ threshold in Figure \ref{fig5}. CSO sources currently below this threshold are Mtype\,1 flaring or Mtype\,3 dying sources. 
  
\subsection{Variation with Kinematic Age} 

\begin{figure}[t]
\includegraphics[scale=0.23,angle=0]{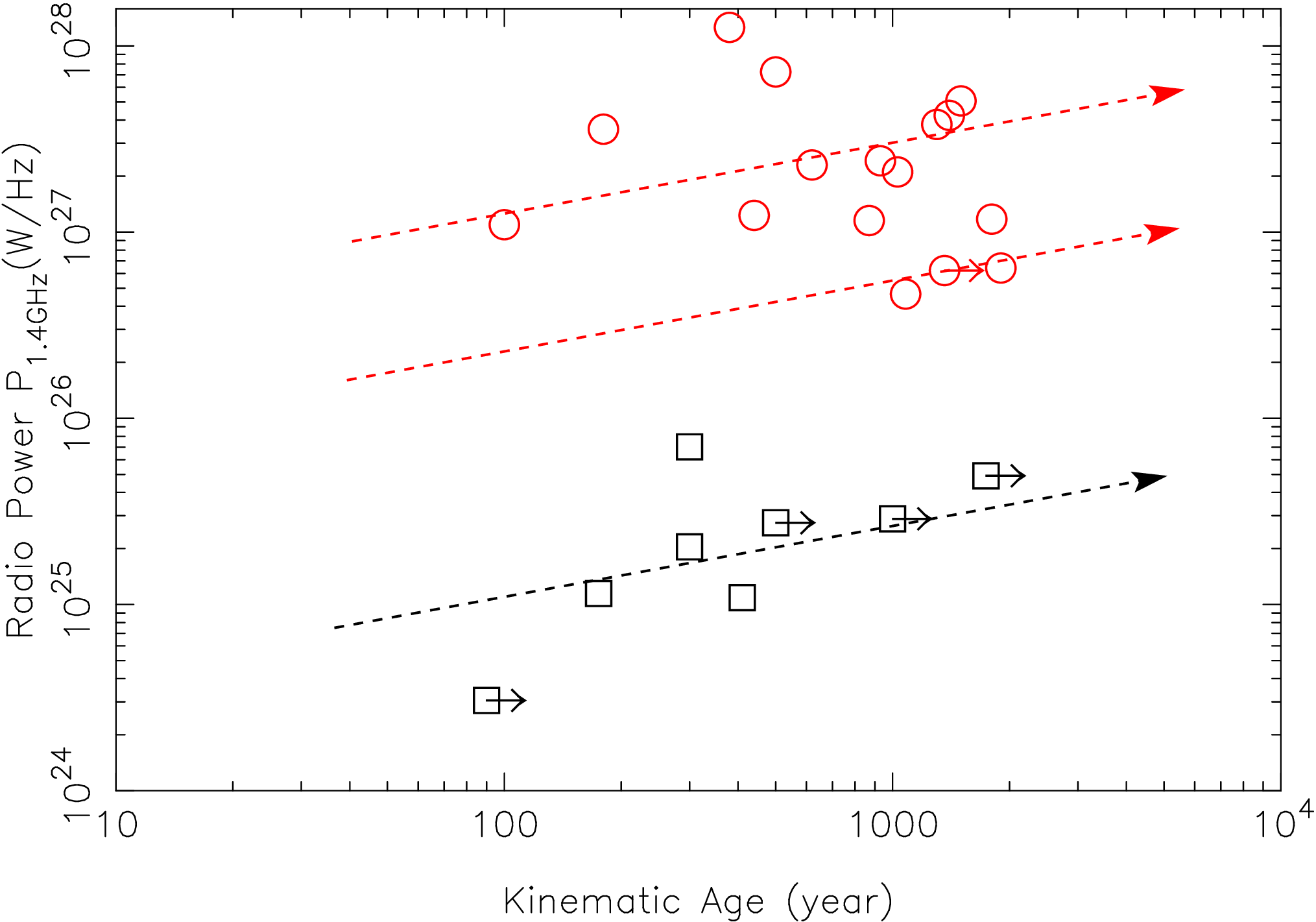}
\caption{Variation of source radio power with kinematic age. Modeling predicts an evolutionary  relation of $P_{rad}\propto T_{kin}^{2/5}$ (dashed arrows). 
\label{fig6}}
\end{figure}

\begin{figure}[t!]
\includegraphics[scale=0.23,angle=0]{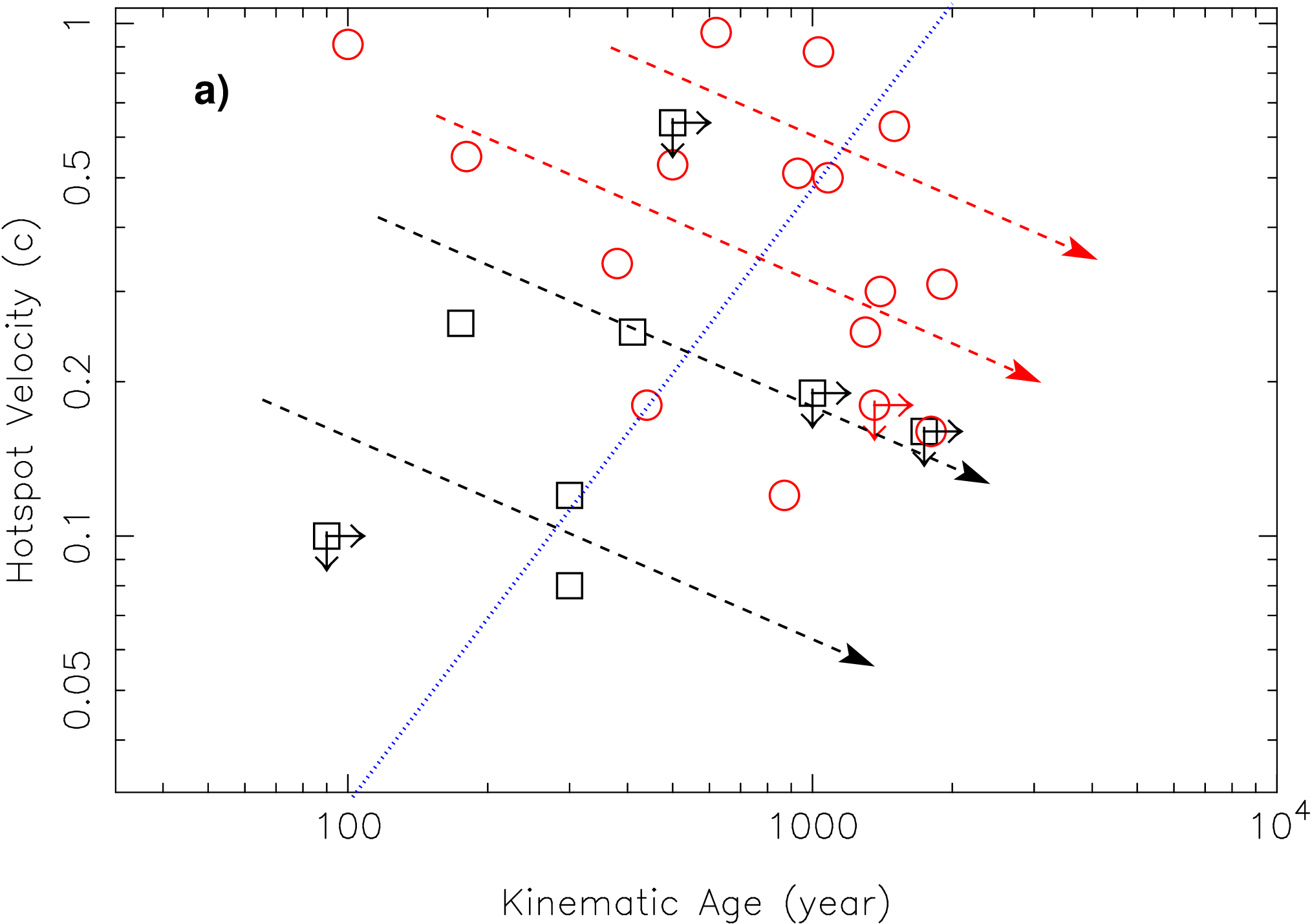}
\includegraphics[scale=0.23,angle=0]{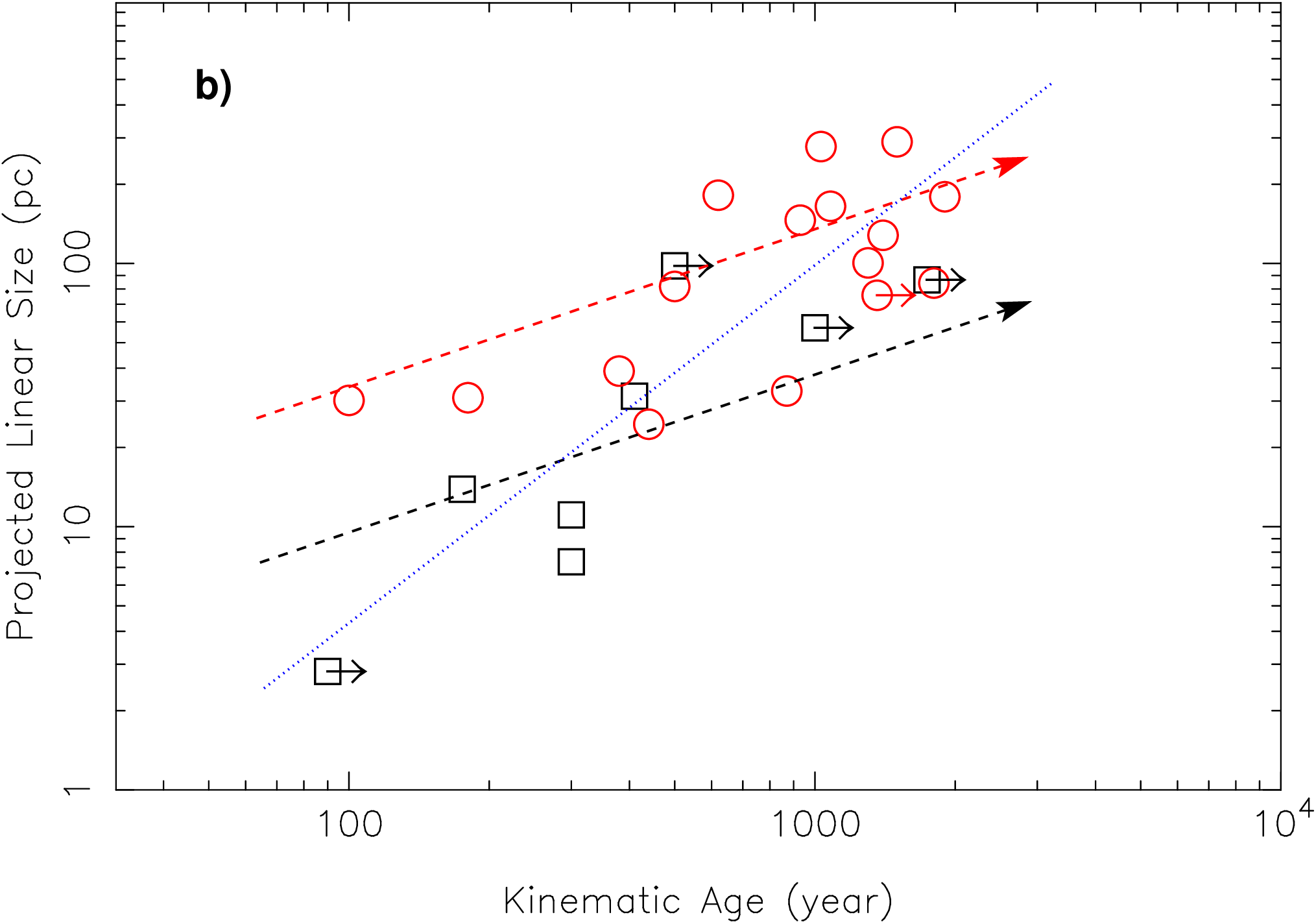}
\caption{Variation of the linear size and the separation velocity with kinematic age. 
(a) Both source groups (high and low radio power) are consistent  with hotspot separation velocity decreasing with age. The predicted evolution of the sources follows a $v_{HS} \propto T_{kin}^{-2/5}$ relation (dashed arrows). (b) The CSO linear size shows an expected increase with age. The predicted variation for individual sources with constant power is $D \propto T_{kin}^{3/5}$ as indicated with dashed arrows. As a group also the older sources have larger sizes as $D \propto T_{kin}^{1.5}$ (dotted line).
 \label{fig7}}
\end{figure}

The hotspot separation velocity, the linear extent, and the radio power of the source all vary strongly with kinematic age (or evolutionary time). Modeling results assuming a constant jet power $P_{src}$ during Phase 1 predict an increase of the radio power with evolutionary time $P_{rad} \propto T_{kin}^{2/5}$ (Sect. \ref{model}). This relation may explain the data points using the grouping in the power levels (Fig. \ref{fig6}). 

The hotspot velocity shows a large spread and are consistent with a decrease with Kinetic Age in Figure \ref{fig7}a.  Modeling predicts a $v_{HS} \propto T_{kin}^{-2/5}$ evolution for a source with fixed power $P_{src}$ and deceleration for the hotspot with time for CSOs.  

The linear size of a CSO naturally increases with the Kinematic Age as is manifested in Figure \ref{fig7}b. Modeling predicts that $D \propto T_{kin}^{3/5}$ for an individual source with constant $P_{src}$ (dashed arrows) (see Sect. \ref{model}). The general trend that older sources have larger sizes is manifested by the whole group and may be expressed with a $D \propto T_{kin}^{1.5}$ as indicated with the dotted line. 
  
 \begin{figure}[t]
\includegraphics[scale=0.23,angle=0]{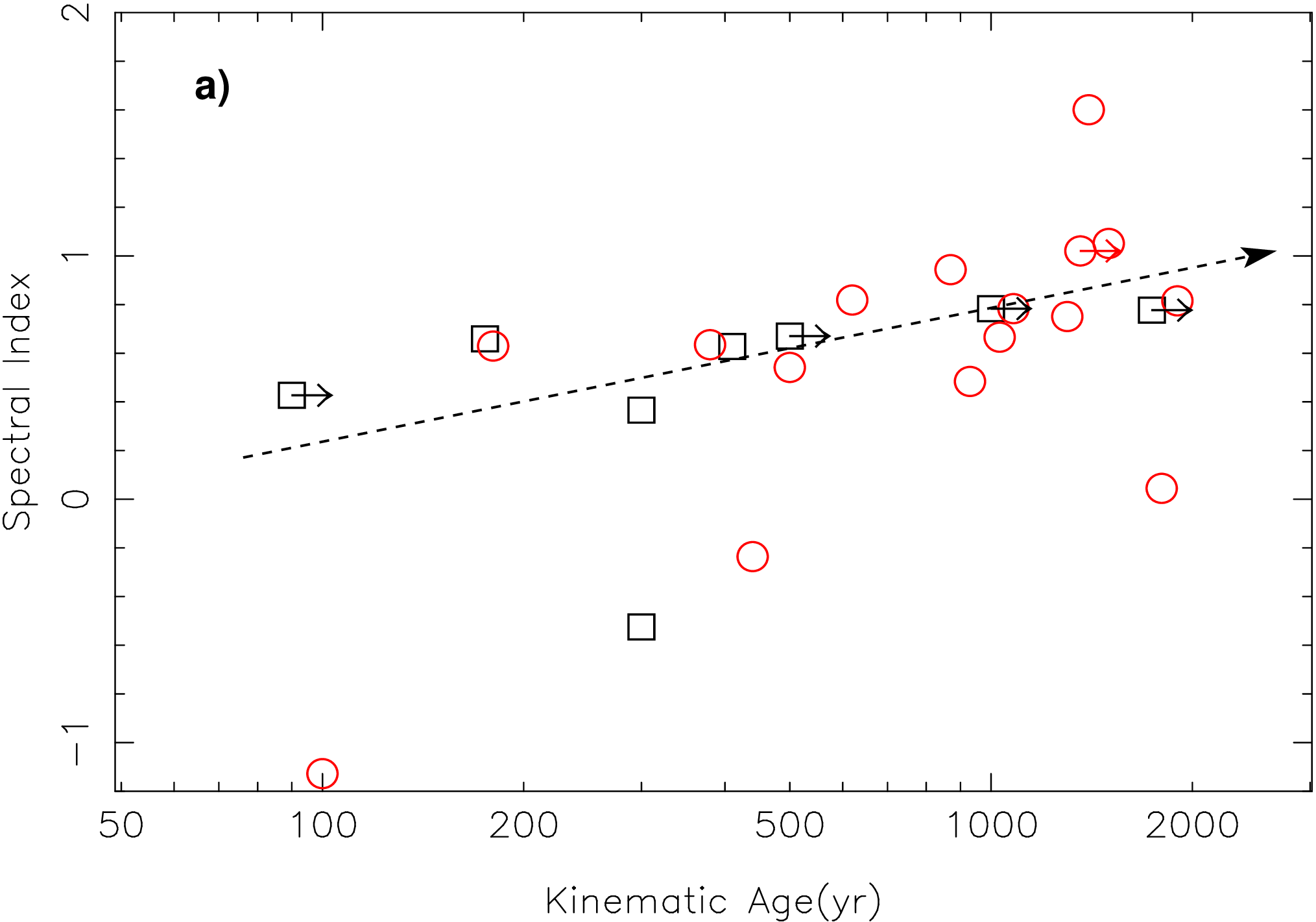}
\includegraphics[scale=0.23,angle=0]{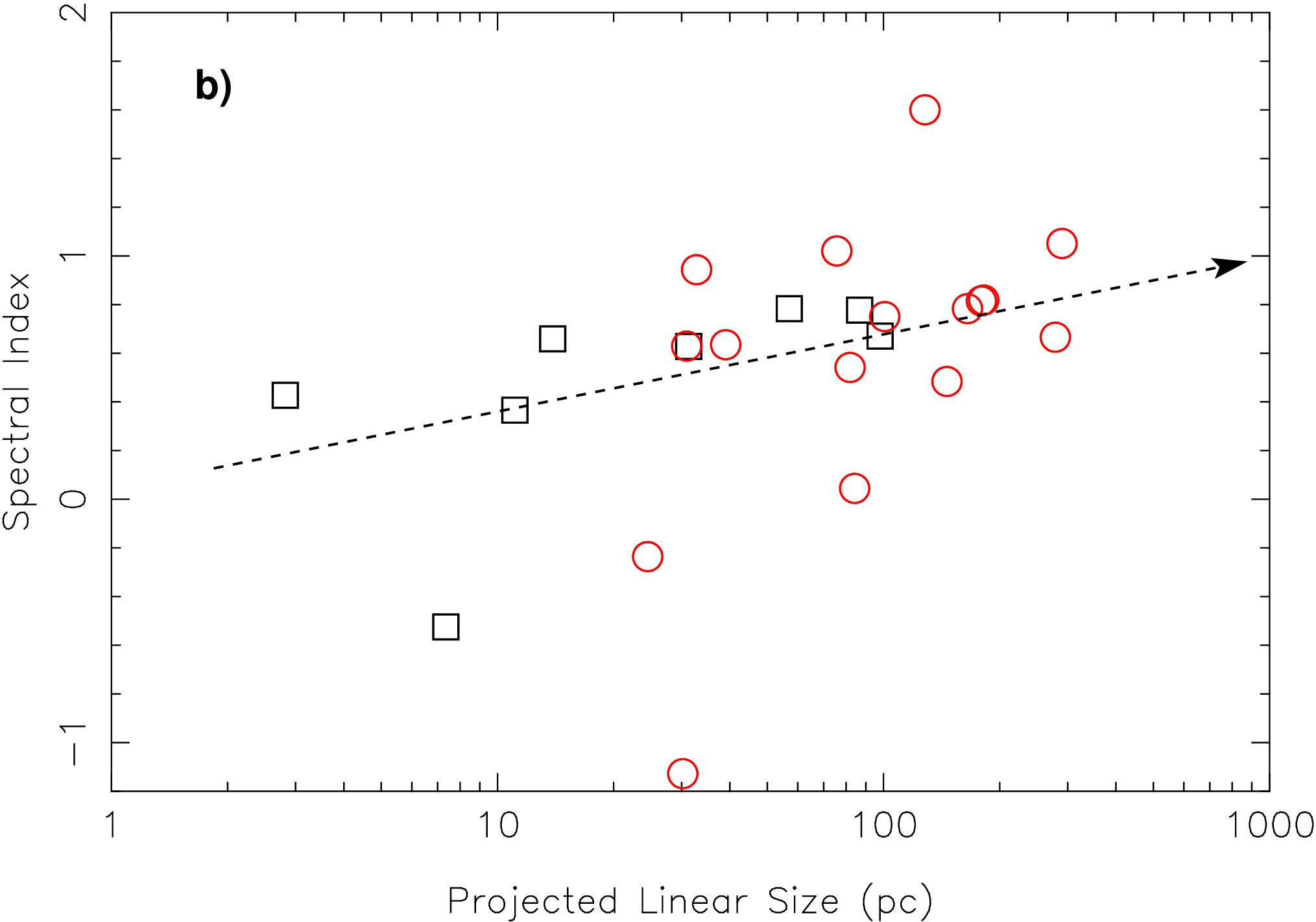}
\caption{Variation of spectral index with evolutionary time and with linear extent. The spectral index between 4.8 and 8.4 GHz varies from $\alpha$ = 0.3 to unity for the range kinematic ages and of linear size. The four HFP sources with inverted spectra have a spectral turnover above 5~GHz. The source with the steepest spectrum is most likely a relic.
\label{fig8}}
\end{figure} 
  
 \subsection{Spectral Index versus Time and Size}  

The spectral index will vary strongly when a CSO evolves from (smaller) adiabatic expansion dominated sources to (larger) synchrotron-loss dominated sources. Synchrotron self-absorption will dominate for these small-size CSO sources. Modeling of the radio properties suggests that the spectral index for CSOs should be close to unity (Sect. \ref{evol}). 

In the diagrams of Figure \ref{fig8}, the spectral index  $\alpha^{8.4}_{4.8}$, defined as $S_\nu \propto \nu^{-\alpha}$, is plotted against kinematic age $T_{kin}$ and source size $D$. The spectrum becomes steeper for larger kinematic ages and linear sizes as the synchrotron self-absorption opacity gradually decreases. Both diagrams display this expected general increase from $\alpha=0.3$ to unity during the CSO stage. Sources that do not follow the trend include a single (relic) source above the distribution and four High Frequency Peakers with a very high turnover frequency ($\nu_{to}>$5 GHz) below the distribution.

The spectral index for the CSO sample presented here is taken in the optically-thin part of the spectrum above the turnover frequency of the sources, which decreases systematically with sources size \citep{ODea98}. The observed range of of spectral indices above the peak for samples of CSS and GPS sources is $\alpha = 0.5 - 0.9$ \citep{ODea90}, which partially overlaps with the CSO range.  
The same range was derived for the spectral index above spectra peak for GPS and FR galaxies, suggesting that particle acceleration and energy loss mechanism preserve the same average spectral index over most of the lifetime of the source \citep{deV09}. Variability of the spectral index has also been associated with the HFP sources and to identify truly young sources \citep{Ori07}.

The observed evolution of the spectral index of the CSO sample confirms the (expected) systematic change as the source gets older and larger, which is explained by the shift of self-absorption to lower frequencies. The youngest CSO and HFP sources have the flattest spectrum. The observed CSO trend should continue into the MSO stage, where it should level off to continue with a ($\alpha = 1$) steep spectrum.

\begin{figure}[t]
\includegraphics[scale=0.32,angle=-90]{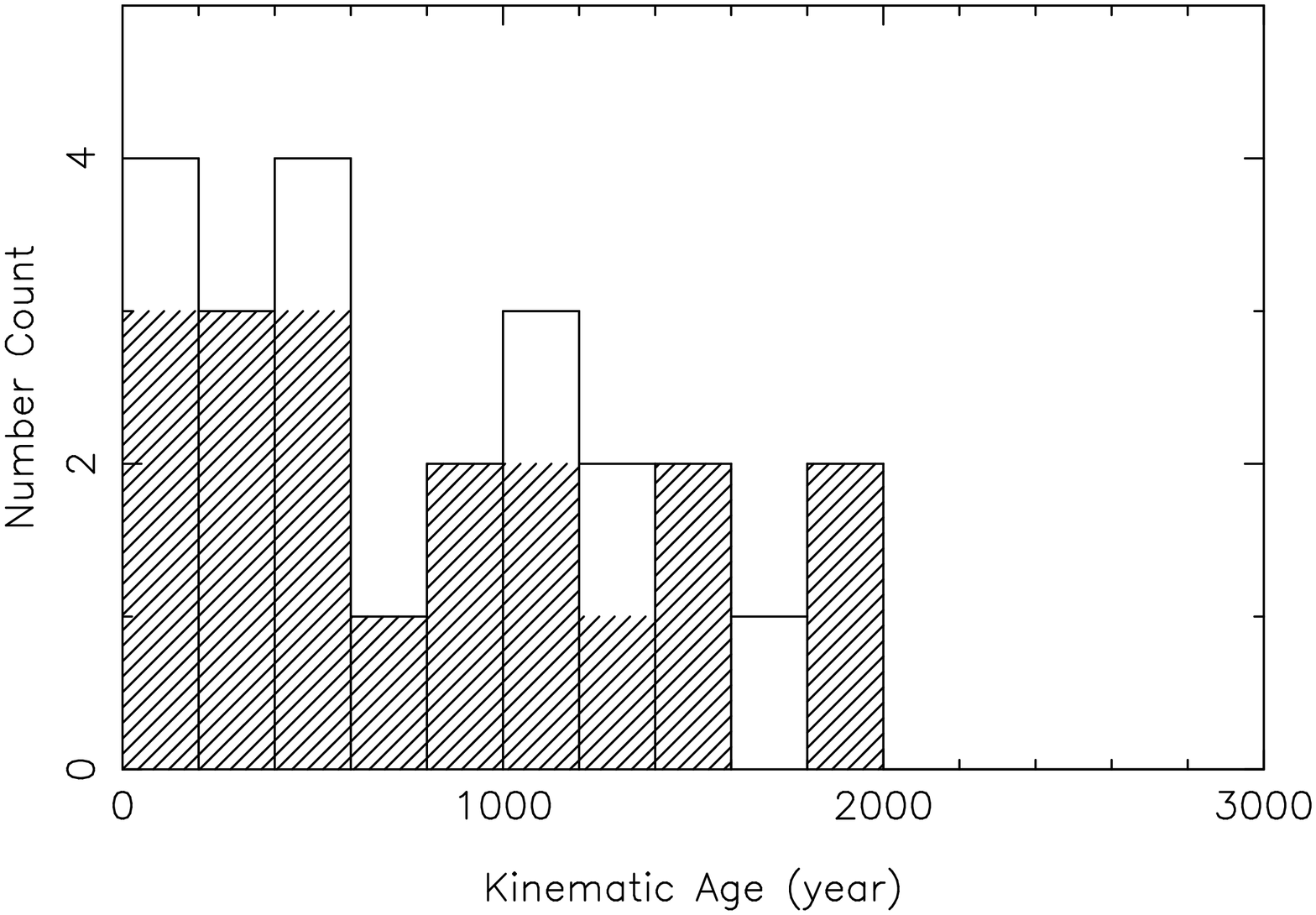}
\caption{Histogram of kinematic ages for the sample of 24 CSOs with known redshifts and expansion velocities 
{\bf from Table 1}. The open squares represent lower limits to the kinematic age. 
\label{fig9}}
\end{figure}

\begin{figure}[t]
\includegraphics[scale=0.32,angle=-90]{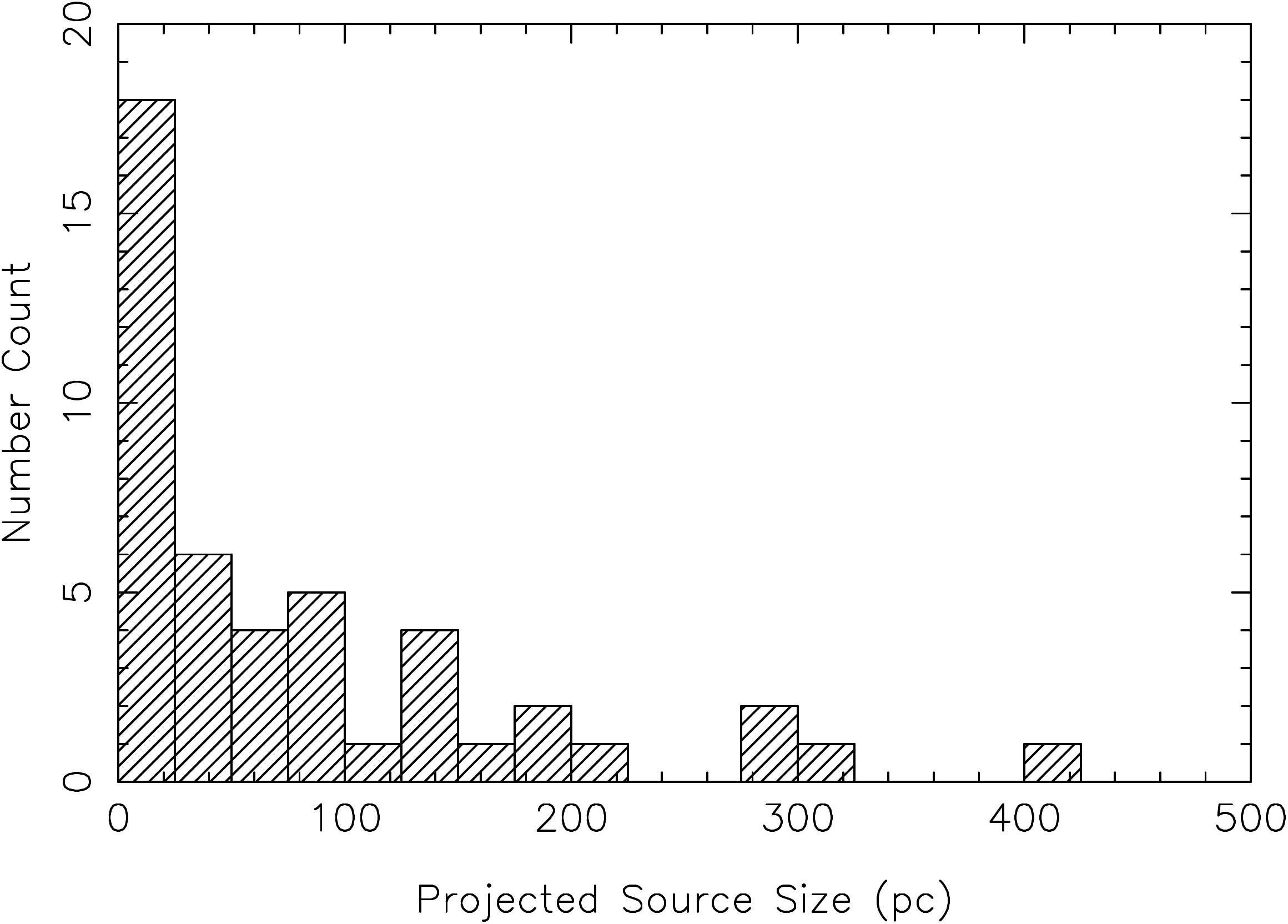}
\caption{Histogram of the projected source size for a larger sample of {\bf 46} CSOs with known redshifts and size information 
{\bf from Table 1}. 
}
\label{fig10}
\end{figure}

\section{Statistics of the CSO sample}

This unbiased sample of CSO sources used in this paper is found not to be uniformly distributed in power, age and size. In general, the CSO phase represents a short episode in the life of a radio source and a uniform distribution of CSO sources of all ages would produce a flat distribution with age.  However, the histogram of the Kinematic Age shows a steadily decreasing distribution (Fig. \ref{fig9}), which suggests that some CSOs will not become old.  The halfwidth of this distribution is on the order of 1100 years. The relative excess of young sources confirms the earlier results of \citep{Gug05,An12}. 

Similarly, the histogram of $D$ shows a decreasing number of sources having larger sizes (Fig. \ref{fig10}).  Projection effects can not be the cause of this distribution because for a randomly oriented sample, the probability for a source to be close to the plane of the sky is much larger than for it having a large inclination angle ({\it i.e}, pointing at the observer). The probability distribution for a distribution with random inclination angles and a given size $D$ increases linearly until it reaches $D$ (a triangular distribution). Similarly, a uniform distribution with random source sizes would also give a flat distribution if all grow larger into CSS sources. The observed $D$ distribution indicates a halfwidth of about 100 pc for the $D$ histogram. The flat distributions at higher $D$ and Age suggest that sources older than typically 1100 years and larger than 100 pc have a better survival rate and will evolve beyond the CSO stage. 

The excess of sources at small sizes and Kinematic Ages indicate that a significant fraction of CSO sources does not continue to evolve beyond the CSO stage towards the MSO stage. This depletion of CSOs may result from reduction/cessation of the jet power and the ensuing instabilities, or from total termination of the nuclear activity after a short activity period. In addition, any recurrent or periodic ejection of emission components would result in the re-excitation of an existing lobe cavity created during earlier outbursts but even such sources may not grow into Mpc-scale doubles. This implication is supported by CSO observations showing sub-pc-scale extended relics, for example,  1511+0518 and OQ~208 \cite{OD10, Wu12}.

\subsection{Transient CSO sources}

{\it Inspection of the source morphology} using the above considerations of a larger sample of 46 CSO sources with high-dynamic-range VLBI images from the literature suggests that only 13 (29\%) sources are Mtype\,2 double sources with well-confined hotspots at the leading edge of the lobes. Similarly within the current sample of 24 sources with kinematic data, a total of 7 (also 29\%) sources have such a well-defined Mtype\,2 signature. The remaining sources show decaying structures with prominent jets, hotspots before the lobes, and diffuse or disrupted lobes. Several sources may even qualify as Mtype\,3 radio relics. 

In principle, the evolutionary characteristics provided in this paper would also allow the identification of the evolutionary state of the CSOs. For instance, the largest sources with the lowest expansion velocities could be precursors of Mtype\,,1 flaring or Mtype\,3 dying sources (in Fig. \ref{fig5}) and a high separation velocity at a high age may indicate survival of the CSO stage. However, the sources in our sample classified as Mtype\,2 doubles do not present any clear tendencies or systematics in the evolutionary diagrams of this paper. Therefore, the current evolutionary data do not (yet) provide discriminatory guidance about the chance of survival of an individual CSO and its morphological classification. 

\subsection{Transient behaviour}

The long-term AGN duty cycle of $10^{7-8}$ yr is generally thought to be induced by major mergers and long-term accretion events. However, the current data suggests that a significant fraction of the CSO population consists of short-lived radio sources with a typical lifetime less than 1000 years.   Similarly, histograms of GPS and CSS source sizes show flat distributions followed by a steep distribution for FR\,I or II sources \citep{ODea98}. Therefore, some GPS and CSS sources will not evolve into classical doubles and instead become transient or frustrated  \citep{ODeaB97}. A strong similarity may thus exist with X-ray/optical/radio transient events in AGN resulting from tidal disruption of stars close to the supermassive black hole (SMBH).

Short-term AGN duty cycles of $\leq10^{4}$yr may result from transient nuclear accretion events or instabilities in the accretion disk. CSO radio power of $10^{27} W/Hz$ (at 5 GHz) can be sustained by SMBH accretion when the accreted mass is:

 $\Delta M = 1.5 \times10^{-3} (\alpha\, \eta)^{-1} \Delta T$ \msols,
 
\noindent where $\Delta T$ is the time interval in years, $\alpha \approx 0.1$ represents the accretion efficiency with half the available mass accreting, and $\eta \approx 0.01$ represents the overall conversion efficiency from accreted energy to radiative energy.
A medium power CSO would thus require about 150 \msol for a 100 year transient episode. Such numbers make source transience more plausible and may account for the disappearance of young CSOs.

\section{Conclusions}

In order to classify the energetics and morphology of CSOs and other radio sources, seven morphological types may be identified that also incorporate the large-scale FR\,I and FR\,II phenomenology. Lower power {\it Mtype\,1 flaring} sources are FR\,1-like sources that show symptoms of jet decay, loss of power, diffuse lobes, or large-scale jet instability.  They exist below the transition region in the $P_{rad}-D$ diagram, unless they are re-energized. Higher power {\it Mtype\,2 double} sources are FR\,II-like sources with efficient (often invisible) jets and well-confined hotspots and lobes. Such sources will follow the complete evolutionary path in the $P_{rad}-D$ diagram until higher $D$. Once an Mtype\,2 source reaches a stage where the jets become unstable, either by growth of instabilities or because of insufficient power, this Mtype\,2 source turns into an Mtype\,1 source. 

Cessation or intermittency of jet power can happen during any evolutionary stage, and a {\it  Mtype\,3 dying} source will move downward in the $P_{rad}-D$ diagram (see Fig. \ref{fig2}).  Therefore, throughout the whole $P_{rad}-D$ diagram there may be Mtype\,3 dying sources as relics of burned-out sources. 

{\it Mtype\,4 restart} sources re-energize Mtype\,3 dying sources or even low-power Mtype\,1 flaring sources. The morphology of the Mtype\,4 restart sources will depend strongly on the state of the sources before the startup. The morphology will differ strongly if the restart occurs in an Mtype\,2 double source or in an Mtype\,3 relic source. X-shaped radio sources would also fit in this morphological type. {\it Mtype\,5 obstructed} sources occur among the CSO and MSO population and show evidence of blockage or bending of the jet flow, an example is the CSS 3C~48 \cite{An10}. {\it Mtype\,6 core-jet} sources are subject to strong projection effects, while host galaxies of {\it Mtype\,7 trail} sources have a significant motion through the intra-cluster medium.

CSOs show a compact-double radio structure that is analogous to classical extended doubles,  although their sizes can be as much as five magnitudes smaller. The morphological similarity among these radio sources invokes an evolutionary scenario from CSO via MSO to LSO.
Their evolution is characterized by the relations between radio luminosity, the separation distance of the two hotspots, the advancing velocity of the hotspots, and the age of the source. The observable properties of small-scale CSOs illustrate the factors that govern their dynamical evolution and eventually their fate, and the information of the physical properties of the ISM of the host galaxies on sub-kpc scales. 
For the purpose of a dynamic evolution study, an unbiased but incomplete sample of 24 CSO sources has been selected solely on the basis of the availability of redshifts and hotspot separation velocities. 

Self-similar evolution models have been used to depict the dynamic evolution of extragalactic Mtype\,2 radio sources in four stages {\it i.e.}, CSO--GPS--CSS--FR\,II. The CSOs are in the earliest growing stage of radio sources where the radio power varies strongly with source size and adiabatic expansion dominates over synchrotron losses inside the ISM-IGM transition point  where the radial dependence ($\beta$) of the ISM density changes. This distance falls in the range of 1$-$3 kpc but it can also be significantly larger. 
The initial CSO sample indeed shows systematic behavior that is consistent with self-similar model predictions \cite[e.g.,][]{KB07}.
 
In all diagnostic diagrams using the CSO observables, the evolutionary path of individual sources can be parameterized by the predictions of self-similar modeling, assuming that the source maintains a constant jet power.  
According to these models an individual CSO increases its luminosity with evolutionary time and with size until it reaches the ISM$-$IGM transition region.  Individual sources have a decreasing separation velocity with increasing power, with evolutionary time, and with size until the ISM$-$IGM transition.  The sources systematically increase in size  with evolutionary time.

The spectral index of the CSO sources changes from zero (flat) to unity (steep) during the course of their evolution until the ISM$-$IGM transition. This aspect is related to the self-absorption during the earliest stages of evolutions and is not covered by self-similarity models.

The CSO sample as a whole is not well-confined, because sources with different jet powers and different environmental parameters are superposed in the diagrams.  Nevertheless, the sample as a whole shows global variation that differs from the evolutionary path of individual sources. More powerful sources are larger and have higher expansion velocities, but cover the same range of kinetic age. Sources with higher expansion velocities are larger and larger sources are also older.

In these models the hotspot separation velocity of a CSO with constant jet power varies as $D^{-2/3}$ within the $\beta = 0$ region. Beyond the ISM-IGM transition the velocity in the jet may increase again if the jet flow remains laminar and supersonic. A sources that do not fulfill these conditions will not have a sustained growth beyond the ISM-IGM transition.  

The histogram of the CSO age estimates shows a lack of old sources, indicating that a significant fraction of CSOs will not evolve into MSOs and eventually LSOs. Sources with current age of $>$1100 years and size of $>100$ pc would have a better chance of survival.  Classification of the current larger CSO sample sample shows that only 29\% of the sample have clear Mtype\,2 double qualities and these are likely to grow further in size.  
 
The dynamic properties and evolutionary behavior of CSOs represent both the {\it youth} and {\it frustration} scenarios. 
CSOs are naturally young radio sources at their early evolutionary stages.  The CSOs with short-lived or intermittent nuclear activity and/or having lower-power can be regarded as frustrated or obstructed and will not evolve into full-size Mtype\,2 double sources.    

Further study of the dynamics of CSO requires a more complete sample of sources, because of the observational bias against the numerous high frequency peakers at the earliest CSO stages, particularly at higher power (compare Fig. \ref{fig3}).  The most compact radio sources are only identified as CSOs when the radio structure is resolved at the VLBI resolution. As a result, only sources with a hint of mas-scale structure from earlier VLBI surveys have warranted follow-up observations. A systematic unbiased VLBI survey of a complete GPS sample would thus be necessary to identify the most compact CSO population.  In addition, current CSO kinematic samples focus more on the high-power population and more sensitive VLBI observations of low-power CSOs are required for a complete view of the CSO evolution.
 
\acknowledgments
The authors thank the anonymous referee for helpful comments which improved the manuscript. 
T.A. thanks the Overseas Research Plan for CAS-Sponsored Scholars, the Netherlands Foundation for Sciences (NWO), and the China$-$Hungary Collaboration and Exchange Program of the CAS. 
This work is supported in part by the 973 program of China (2009CB24900, 2012CB821800), the KNAW-CAS exchange program (code: 10CDP005), the Science \& Technology Commission of Shanghai Municipality (06DZ22101) and the  Strategic Priority Research Program of the CAS (XDA04060700).
This research has made use of the NASA/IPAC Extragalactic Database (NED), which is operated by the Jet Propulsion Laboratory, California Institute of Technology, under contract with the National Aeronautics and Space Administration.

\input{table1.tex}

\end{document}

%% file: table1.tex

\begin{deluxetable}{ccccccccccccccccc}
\tablecolumns{16}
\tabletypesize{\scriptsize}
\setlength{\tabcolsep}{0.03in}
\tablewidth{0pt}
\tablecaption{Radio Properties of a sample of {\bf 46} CSOs and Candidates}  
\renewcommand{\arraystretch}{1.0}
\tablehead{
Source &
$z$    & 
$D_L$  & 
$\theta_{AS}$ & 
LS     &
$\mu$  &
$v$   &
Age   &
ID    &
Ref.  &
$S_{1.4}$&
$S_{4.8}$&
$S_{8.4}$& 
$\nu_{to}$ &
$\alpha^{8.4}_{4.8}$& 
$log_{10}P_{1.4}$   &
Ref.   \\
      &      
      & 
(Gpc) & 
(mas) & 
(pc)  &
($\mu$as/yr)&
($c$) &
(yr)  &
($^{a}$)      & 
(Map) &
(Jy)  &
(Jy)  &
(Jy)  &
(GHz) &
($^b$)& 
(W/Hz) &  \\
(1) & (2) & (3) & (4) & (5) & (6) & (7) & (8) & (9) & (10) & (11) & (12) & (13) & (14) & (15) & (16) & (17)
} 
\startdata
   0003+2129  &0.455 &2.519  & 3.8  &21.9 &            &     &             &      &O06      &0.100 &0.260 &0.227 & 5.4     &0.23(0.9)   &26.92    &O07        \\
   0005+0524  &1.887 &14.641 & 1.7  &14.5 &            &     &             &      &O06      &0.166 &0.210 &0.165 & 3.7     &0.43(0.6)   &27.91    &O07        \\
   0029+3456  &0.517 & 2.854 & 32.0 &200.0&            &     &             &      &K04      &1.750 &1.312 &0.981 &0.4--1.4 &0.52        &27.02    &W92,G91,H07\\
   0038+2302  &0.096 &0.419  &18.5  &31.3 &41.0        &0.25 & 410         &Cl\,I &P09      &0.532 &0.247 &      &0.4--1.4 &0.62        &25.03    &W92,G91    \\     
   0048+3157  &0.015 &0.0582 & 4.4  &1.2  &            &     &             &      &VCS      &0.270 &0.802 &0.286 &1.4--5.0 &1.88        &24.53    &W92,G06,H07\\  
   0111+3906  &0.668 &3.917  & 5.8  &39.0 & 9.3        &0.34 & 380         &Cl\,I &OC98     &0.509 &1.301 &0.918 & 4.7     &0.62(1.1)   &28.10    &O07        \\     
   0119+3210  &0.060 &0.255  &89.4  &98.0 &$<$170.0 &$<$0.64 &$>$500       &Cl\,I &G03      &2.826 &1.571 &1.087 & 0.4     &0.67        &25.44    &W92,G91,H07\\ 
   0428+3259  &0.479 &2.680  & 2.7  &16.0 &            &     &             &      &O06      &0.148 &0.524 &0.531 & 6.9     &$-0.02$(0.6)&27.06    &O07        \\
   0431+2037  &0.219 &1.045  &42.7  &145.6&            &     &             &      &F00      &3.611 &2.300 &1.710 & 0.63    &0.54        &26.73    &W92,G91,H07\\
   0503+0203  &0.584 &3.322  &10.36 & 66.5&            &     &             &      &S01      &2.118 &2.197 &1.411 & 2.5     &0.81        &27.59    &W92,G91,H07\\
   0650+6001  &0.455 &2.519  & 3.2  &18.5 &            &     &             &      &O06      &0.507 &1.150 &0.975 & 5.3     &0.29(0.5)   &27.18    &O07        \\
   0713+4349  &0.518 &2.870  &24.3  &146.0&17.0        &0.51 & 930         &Cl\,I &PC03     &1.940 &1.670 &1.280 & 1.9     &0.48        &27.38    &S98        \\     
   0943+5113  &0.42  &2.289  & 3.9  &21.5 &            &     &             &Cl\,II&OD12     &0.077 &0.147 &0.064 & 3.7     &1.48(1.8)   &26.10    &OD12,ODS10 \\
   0951+3451  &0.29  &1.482  & 4.8  &20.7 &            &     &             &Cl\,II&OD12     &0.019 &0.062 &0.055 & 6.0     &0.21(0.6)   &25.66    &OD12,ODS10 \\
   1035+5628  &0.450 &2.483  &32.2  &181.8& 37.6       &0.96 & 620         &Cl\,I &T96      &1.780 &1.270 &0.810 & 1.3     &0.81        &27.36    &S98        \\     
   1111+1955  &0.299 &1.493  &17.7  & 75.7&$<$10.0  &$<$0.18 &$>$1360      &Cl\,II&G05      &1.152 &0.648 &0.370 &0.4--1.4 &1.02        &26.79    &W92,G91,H07\\
   1120+1420  &0.362 &1.875  &85.06 &415.9&            &     &             &      &S95      &2.460 &1.000 &0.613 & 0.5     &0.8         &27.00    &S98        \\
   1148+5254  &1.632 &12.228 & 0.8  & 4.3 &            &     &             &      &O06      &0.108 &0.414 &0.450 & 7.9     &$-$0.15(0.4)&27.96    &O07        \\
   1247+6723  &0.107 &0.479  & 7.4  & 13.9&38.0        &0.26 & 175         &Cl\,II&P09      &0.344 &0.191 &0.133 &0.4--1.4 &0.65        &25.06    &W92,G91,P92\\     
   1256+5652  &0.042 &0.181  & 70.0 & 56.0&            &     &             &      &VCS      &0.288 &0.419 &0.257 &$<$0.07  &0.89        &24.69    &W92,G91,H07\\
   1309+4047  &2.91  &24.905 & 0.8  & 6.3 &            &     &             &Cl\,I &OD12     &0.037 &0.131 &0.115 & 5.4     &0.23(0.8)   &28.22    &OD12       \\
   1324+4048  &0.496 &2.719  & 5.6  & 32.7& 4.2        &0.12 & 870         &Cl\,II&A12      &0.357 &0.413 &0.246 &1.4--4.9 &0.94        &27.06    &W92,G91,P92\\     
   1326+3154  &0.368 &1.910  &56.4  &278.1& 40.0       &0.88 &1030         &Cl\,II&K98      &4.631 &2.350 &1.630 & 0.5     &0.66        &27.32    &S98        \\     
   1335+4542  &2.449 &20.183 & 1.3  & 4.1 &            &     &             &      &O06      &      &0.821 &0.646 & 5.0     &0.43(0.7)   &28.88    &O07        \\
   1335+5844  &0.57  &3.31   &13.3  & 84.2&  4.7       &0.16 &1800         &Cl\,I &A12      &0.299 &0.744 &0.726 & 6.0     &0.04(0.3)   &27.07    &O07        \\
   1400+6210  &0.431 & 2.300 &57.6  &313.3&            &     &             &      &D95      &4.490 &1.800 &1.200 & 0.5     &0.7         &27.41    &S98        \\
   1407+2827  &0.0766&0.333  & 8.0  & 11.1&25.0        &0.13 & 255         &Cl\,I &W12      &0.865 &2.532 &2.071 & 5.1     &0.36(1.0)   &25.85    &O07        \\     
   1414+4554  &0.186 &0.874  &28.8  & 86.7&$<$14.0  &$<$0.16 &$>$1740      &Cl\,I &G05      &0.360 &0.213 &0.139 &$\sim$1.4&0.78        &25.69    &W92,G91,P92\\     
   1415+1320  &0.247 &1.202  & 8.1  & 30.2&60.0        &0.91 & 100         &CD    &G05      &1.206 &0.842 &1.564 &8.4--15.0&$-$1.11(0.6)&27.04$^c$&W92,G91,H07\\ 
   1511+0518  &0.084 &0.370  & 4.8  &  7.3&14.0        &0.08 & 300         &Cl\,I &A12      &0.092 &0.608 &0.811 &10.7     &$-$0.51(0.3)&25.31    &O07        \\     
   1546+0026  &0.550 & 3.087 & 6.0  &37.4 &            &     &             &      &G05      &1.808 &1.149 &0.889 &$<$1.4   &0.47        &27.27    &W92,G91,H07\\
   1559+5924  &0.0602& 0.259 & 7.5  & 8.4 &            &     &             &      &B04      &0.229 &0.197 &0.130 &0.4--1.4 &0.76        &24.59    &W92,G91,H07\\
   1609+2641  &0.473 &2.571  &50.6  &290.0&23.0        &0.63 &1500         &Cl\,I &N06      &4.860 &1.710 &0.960 & 1.0     &1.05        &27.71    &S98        \\     
   1616+0459  &3.197 &27.908 & 1.4  &10.8 &            &     &             &      &O06      &0.317 &0.771 &0.559 & 5.0     &0.57(1.0)   &29.49    &D00,T05    \\
   1723$-$6500&0.0144&0.060  &10.0  &  2.8&$<$106   &$<$0.10 &$>$90        &Cl\,II&PC03     &3.540 &4.640 &3.670 &1.4--2.5 &0.40        &24.48    &T03        \\     
   1734+0926  &0.735 &4.402  &14.2  &100.5&  6.3       &0.25 &1300         &Cl\,II&A12      &1.110 &0.740 &0.490 & 2.3     &0.75        &27.58    &S98        \\     
   1755+6236  &0.027 &0.1148 & 8.3  & 4.4 &            &     &             &      &B04      &0.288 &0.198 &0.147 &$<$0.3   &0.54        &23.78    &C98,G91,H07\\
   1816+3457  &0.245 &1.185  &36.0  &135.0&            &     &             &      &G05      &0.678 &0.355 &0.217 &$\sim$0.4&0.90        &26.25    &W92,G05,G05\\
   1823+7938  &0.224 &1.072  &15.8  &54.8 &            &     &             &      &VCS      &0.297 &      &0.592 &$\sim$8.4&(1.0)       &26.65    &W92,G05,G05\\
   1845+3541  &0.764 &4.615  & 5.6  & 31.0& 13.0       &0.55 & 180         & Cl\,I&PC03     &1.031 &0.794 &0.562 &1.4--4.9 &0.62        &27.55    &W92,G91,P92\\     
   1939$-$6342&0.181 &0.846  &43.4  &128.0&26.0        &0.30 &1400         &Cl\,II&GP09     &14.98 &5.840 &2.425 &$\sim$1.4&1.62        &27.63    &T03,H07    \\     
   1944+5448  &0.263 &1.285  &42.2  &165.0&31.0        &0.50 &1080         &Cl\,I &PC03     &1.647 &0.938 &0.610 &0.4--1.4 &0.78        &26.67    &W92,G91,P92\\     
   1945+7055  &0.101 &0.445  &32.0  & 57.0&$<$30    &$<$0.19 &$>$1000      &Cl\,I &PC03     &0.921 &0.645 &0.477 &0.4--1.4 &0.54        &25.46    &W92,G91,P92\\     
   2022+6136  &0.227 &1.086  & 7.0  & 24.5&13.0        &0.18 & 440         & Cl\,I&PC03     &2.110 &2.820 &3.210 & 8.4     &$-$0.23(0.6)&27.09$^b$&S98        \\    
   2203+1007  &1.005 &6.488  &10.5  & 81.8&10.3        &0.53 & 500         & Cl\,I&A12      &0.107 &0.311 &0.231 & 4.8     &0.53(1.0)   &27.86    &O07        \\     
   2355+4950  &0.238 &1.144  &49.3  &179.2&21.1        &0.31 &1900         &Cl\,I &OCP99    &2.341 &1.552 &0.992 & 0.7     &0.81        &26.81    &W92,G91,P92,S98 
\enddata \\
\flushleft
$*$: Sources have both redshift and kinematic data and are used in Figures 3--10; \\ 
$^a$: Source structure identifications are: Class\,I and disturbed structure, Class\,II structure, and CD for core-dominated. One source has also been classified as a High-Frequency-Peaker; \\
$^b$: For CSOs with a spectral peak higher than 4.8 GHz, the spectral index is calculated from the optically thin part of the spectrum above the turnover frequency and given in bracket; \\
$^c$: The 1.4~GHz radio power is extrapolated from the 8.4-GHz flux density and the spectral index $\alpha^{15GHz}_{8.4GHz}$. \\
The references are : A12: \cite{An12}; B04: \cite{Bon04}; D95: \cite{Dal95}; F00: \cite{Fom00}; G91: \cite{GC91}; G03: \cite{Gir03}; G05 \cite{Gug05}; GP09: \cite{GP09}; K98: \cite{Kel98}; K04: \cite{Kel04}; L00: \cite{Liu00}; N06: \cite{Nagai06}; O06: \cite{Ori06}; O07: \cite{Ori07}; OC98: \cite{OC98}; OCP99: \cite{OCP99}; OD12: \cite{OD12}; ODC10: \cite{ODC10} P09: \cite{Pol09}; P92: \cite{Pat92}; PC03: \cite{PC03}; PT00: \cite{PT00}; S95: \cite{San95}; S98: \cite{Sta98}; S01: \cite{Sta01}; T96: \cite{TRP96}; T03: \cite{TK03}; VCS: VLBA Calibrator Survey \cite{VCS6}, http://astrogeo.org/vcs/; W12: \cite{Wu12}; W92: \cite{WB92}. 
\end{deluxetable}